\makeatletter
\disable@package@load{algorithm2e}{}
\makeatother

\documentclass[pmlr]{jmlr}

\usepackage{booktabs}
\usepackage{float}
\usepackage{algorithm}
\usepackage{algpseudocode}
\usepackage{pgfplots}
\usepackage{siunitx}
\usepackage{tikz}
\usepackage{xurl}
\usepackage{microtype}
\usepackage{bbold}

\pgfplotsset{compat=1.18}
\usepgfplotslibrary{groupplots,statistics}
\usetikzlibrary{arrows.meta,positioning,fit}
\hypersetup{hypertexnames=false}

\jmlrvolume{349}
\jmlryear{2026}
\jmlrworkshop{Conference on Applied Machine Learning for Information Security}

\title[Compile Timestamps and First-Seen Times]{Trouble With Time: Compile Timestamps and First-Seen Date Regression in EMBER2024}

\author{\Name{Edward Raff}
 \Email{edward.raff@crowdstrike.com}  \\ 
 \addr CrowdStrike
 \AND
\Name{Aditya Kapoor}  \Email{aditya.kapoor@crowdstrike.com} \\
\addr CrowdStrike}

\editor{Edward Raff and Ethan M. Rudd}

\begin{document}

\maketitle

\begin{abstract}
Reliable first-seen dates are important for constructing temporal malware evaluation splits, but this information often comes from external services that are not equally accessible to all researchers. We investigate whether machine learning can instead estimate an executable's first-seen date or determine when its PE/COFF compilation timestamp is unreliable. Using EMBER2024, we evaluate feature-based and raw-byte models across classification, regression, ranking, and latent-reliability formulations. Direct first-seen regression produces errors comparable to a substantial portion of the dataset's collection window, and none of the evaluated scores provides a useful chronological ordering of files. Current models therefore cannot replace externally sourced first-seen metadata. In contrast, classifiers can identify timestamps that fall within specified tolerances of first seen and can detect clearly implausible timestamps with high discrimination. We also find substantial timestamp disagreement among both benign and malicious executables, indicating that unreliable timestamps are not exclusively a product of malicious manipulation. These results establish baselines and identify timestamp reliability as a distinct direction for future security-ML research.
\end{abstract}

\begin{keywords}
timestamps; compilation dates
\end{keywords}

\section{Introduction}
\label{sec:introduction}

It is desirable to know when a specific malware sample was produced. The most obvious factor is that it allows us to organize datasets by time, so that we can create train/test splits by a single point in time and thus reduce model evaluation bias~\citep{joyce_maldict_2023}. This information is stored in the metadata of the executable. 
Microsoft’s PE/COFF specification defines the COFF TimeDateStamp as seconds since 1970 \citep{microsoft_pe_format}. But nothing forces this timestamp to be accurate. It is common wisdom within the industry that malware authors will perturb timestamps to confuse or evade investigations. 

It is more reliable to use a ``first submission date'' from services like VirusTotal, which provides a censored view of the sample's true compilation date. This is useful but imperfect; one does not know how much time occurred between the sample's compilation and observation. From our perspective as industry players, we have cases of samples we know have existed for months or years and have not entered any such online service. This partial observability is a natural problem that all industry participants face for a variety of competitive, technical, and legal reasons. But the issue is exacerbated for academic researchers, who have less visibility and may lack access to such services~\citep{raff_survey_2020}. 

In this work, we seek to transcribe several points of ``common knowledge'' from practitioners as explicit research questions, and investigate them and potential remediation. In particular we tackle the following research questions (RQs):
\begin{enumerate}
    \item How reliable are PE/COFF timestamps for usage in benign/malicious programs? 
    \item Can we identify inaccurate timestamps?
    \item Can machine learning alleviate issues in timestamp infidelity by providing a better estimate/correction? 
    \item Are there any indicators of inaccurate timestamps?
\end{enumerate}

We tackle these questions by leveraging the EMBER24 dataset~\citep{joyce_ember2024_2025}. In doing so, we discover a number of unexpected facets and difficulties, to wit, not previously documented in the academic literature. Timestamp variation issues are essentially equally prevalent in benign programs as in malicious ones, which calls into question whether intentional timestamp alteration is truly a malicious activity or an artifact of software engineering practices. We also find that, while one can make reasonable models for ``is this compilation date stale'' relative to first-seen data, multiple strategies to regress the first-seen date itself are largely ineffective. Lastly, there is at least a sub-population of easily identifiable ``bad'' compilation timestamps which we can predict accurately, and we find evidence that benign and malicious programs do exhibit artifacts that signal a potential difference in how benign/malicious samples receive their labels. 

Given the lack of prior research on this problem, we will review the literature on compilation timestamps in \autoref{sec:related-work} and note explicitly that our own work is not the end of this matter, but reveals multiple avenues for future exploration. In \autoref{sec:methods}, we will review the two types of models we build (trees and neural networks), covering six different loss/scenario types we test to understand the broad landscape of approaches one could take and which seem most promising. Exploration of the timestamps themselves, and the results of regressing them, is reviewed in \autoref{sec:results}. These lead us to test an ``obviously bad'' compilation-date model in \autoref{sec:obviously-bad-shap}, which reveals additional insights into the problem, with further analysis to check for potential confounders. Finally, we conclude in \autoref{sec:conclusion}.

\section{Related Work}
\label{sec:related-work}

A large motivation for our work is to understand how far machine learning models can enable effective use of timestamps outside of having first-observed data. The need for this information in academic evaluation was clearly established by ~\citep{235493}, with further analysis by ~\citep{chow_beyond_2026} who first showed that information leakage from randomly mixing samples over time can cause inflated accuracy estimates. Such first-seen information is generally obtained from VirusTotal reports, which are not available to all researchers. Still, such information has been used for analyzing the reports themselves~\citep{joyce_avscan2vec_2023}, and studying how AntiVirus product correlations change in time~\citep{Joyce2021}. Along with many works performing research on malware detection~\citep{raff_peheader_2017,raff_zipf-gramming_2026,angioni_robust_2022,daoudi_lessons_2021,kan_investigating_2021,alam_adapt_2025,liu_explainable_2022,barbero_transcending_2022,li_revisiting_2025,chow_drift_2023,Harang2020}.

VirusTotal has been predominant in the academic literature thus far, and while samples can appear in VirusTotal in hours after being first seen on a device~\citep{van_liebergen_deep_2022}, we are aware of no study on how long it takes samples to appear on an instrumented/observed device, such that they will be seen and captured. Indeed, this is a question involving significant epistemic uncertainty in market penetration, customer differences, and the lag time from executable development to distribution. Such issues often confound analysis in this space and can be irreducible~\citep{agtr}. Historical work indeed shows that other AntiVirus vendors may see a sample 100 days before it is observed by VirusTotal~\citep{berlin_improving_2016}. Similarly, it is likely that VirusTotal also receives samples sooner than the same companies would observe them if it were not for VirusTotal providing a data feed. The important factor is the recognition that the first-seen metadata provided by VirusTotal is an upper bound on existence, and a large predated timestamp may indeed be valid and truthful. 

Timestamp manipulation in PE binaries is a foundational element of adversary tradecraft. MITRE ATT\&CK documents this behavior under T1070.006 (Indicator Removal: Timestomp), cataloguing its use across more than a dozen named threat groups and over forty malware families spanning nation-state actors from Russia, China, North Korea, and Iran, as well as ransomware operators such as BitPaymer~\citep{frankoff_bitpaymer_2018} and BlackByte 2.0~\citep{microsoft_blackbyte_2023}. These techniques have been employed since the early 2000s or earlier~\citep{mitre_t1070_006}.

The motivations for manipulation are varied. First, implausible or malformed \texttt{TimeDateStamp} values in the \texttt{IMAGE\_FILE\_HEADER} can create blind spots in automated detection pipelines. InvisiMole zeroed out PE timestamps entirely~\citep{hromcova_invisimole_2018}, and Misdat's use of Borland Delphi mangled them as a toolchain side effect~\citep{gross_dust_storm_2016}. Second, a valid compile timestamp reveals development timelines, campaign duration, and shared codebases; falsifying it denies this intelligence to defenders. Kimsuky~\citep{dahan_kimsuky_2020}, Gazer~\citep{eset_gazer_2017}, BOOKWORM, and Mustang Panda~\citep{falcone_bookworm_2025} have all deliberately falsified compile timestamps or manipulated PE metadata to obscure module creation times. Third, and most pervasive, is blending into the local file environment. APT29 matched web shell timestamps to neighboring files in the same directory~\citep{mandiant_unc3524_2022}; Lazarus Group cloned timestamps from \texttt{calc.exe} and \texttt{mspaint.exe}~\citep{novetta_blockbuster_2016}; NightClub mimicked \texttt{user32.dll}~\citep{faou_moustachedbouncer_2023}. The goal is to make malicious files appear legitimate and survive triage.

Additional motivations include defeating sandbox and emulator environments---CHIMNEYSWEEP backdated its executable to dates between 2010 and 2021 to manipulate how dynamic analysis environments contextualize the sample~\citep{mandiant_roadsweep_2022}---and corrupting the forensic timeline. APT28~\citep{alperovitch_bears_2016}, APT38~\citep{cisa_fastcash_2020}, Chimera~\citep{jansen_chimera_2021}, and UNC3886~\citep{marvi_esxi_2023} have all used timestomping to obscure breach timelines. POSHSPY backdated downloaded executables to pre-2013 timestamps to exploit age-based detection assumptions~\citep{dunwoody_poshspy_2017}, and UPSTYLE restored original timestamps after acting, erasing evidence of the tampering itself~\citep{volexity_upstyle_2024}. Finally, not all manipulation is intentional: parasitic file infectors sometimes overwrite PE header fields including \texttt{TimeDateStamp} as a side effect of the infection routine. Distinguishing incidental corruption from deliberate falsification is itself a meaningful analytical task.

However, benign instances of unusual or apparently unreliable timestamps have been recognized~\citep{vanini_understanding_2025}. One of the most common offered explanations is the use of reproducible build artifacts, where the timestamp is replaced with a specific marker \texttt{IMAGE\_DEBUG\_TYPE\_REPRO} so that the MD5/SHA hash of a file remains consistent when building the same file at different times~\citep{lamb_reproducible_2022}. This can also occur on the same motivation when build servers are used without intentionally setting any particular value for the timestamp, but the virtual machine used for compilation has an otherwise arbitrary previous timestamp that is reused whenever the virtual machine is restored for another compilation. 

\section{Methods}
\label{sec:methods}

The analysis uses the PE and .NET feature records from EMBER2024. The file first-seen time is taken from \texttt{first\_submission\_date}. The compile time is taken from the PE COFF timestamp field, \texttt{header.coff.timestamp}. For each file, the signed lag is
\[
  \Delta_{\mathrm{days}} =
  \frac{\texttt{first\_submission\_date} - \texttt{header.coff.timestamp}}{86400}.
\]
Positive values mean that the compile timestamp is before first seen. Negative values mean that the compile timestamp is after first seen and is not chronologically plausible. 

The initial distributional analysis used 2,884,868 PE/.NET rows. The rows consisted of 1,440,000 benign files and 1,444,868 malicious files. Missing compile timestamps were excluded from quantile summaries. Results are reported using quantiles because the lag distribution contains future-dated, non-positive, and very old sentinel-like timestamp values.

For predictive experiments, the official EMBER2024 PE feature extractor was used to produce a 2,568-dimensional feature vector. Because several prediction targets are defined directly from the COFF timestamp, we ablate the impact of direct access to that header field by removing the feature at index 696 before training. This leaves 2,567 features and defines the no-COFF condition. This ablation does not remove every feature related to the COFF timestamp: the remaining standard EMBER2024 representation includes Authenticode and cross-field timing information that may provide indirect evidence about it. We retain those features because they also encode distinct and potentially valuable executable metadata; removing them would conflate the impact of the COFF header field with a broader feature ablation. A second condition restores direct access by appending \(\texttt{header.coff.timestamp}/86400\), in days since the Unix epoch, as the final column of the timestamp-removed matrix. The comparison therefore measures the impact of direct access to the COFF header timestamp, rather than attempting to eliminate all related temporal information.

Models were trained on the EMBER2024 train split with 2,339,221 PE/.NET rows. The regular test split contains 539,869 rows. The challenge split contains 4,868 PE/.NET rows, of which 605 have first-seen dates after the maximum train first-seen date. All tables that report challenge-split results are restricted to those 605 post-train challenge rows. Challenge rows overlapping the training time range are used only in time-indexed plots that explicitly mark the train/test time boundary.

\subsection{Modeling}

We will use two sets of models and multiple loss functions to explore what does and does not relate to the ability to make predictions about timestamps, predominantly assuming that the first-seen tags in the data constitute the ground truth regarding the file's ``existence'' date. We will use the original EMBER2024 features with an XGBoost model~\citep{xgboost}, as such domain-knowledge features tend to perform best in malware detection. Since we are not performing malware detection, though, it is possible that the features are not well-designed or informative to the current task. A second MalConv-GCG model (or ``MalConv'' for short)~\citep{Raff2020b,MalConv} is trained from scratch from the raw bytes of the executable. 

All tree models use the EMBER2024-style tree configuration: 500 boosting rounds, histogram tree method, maximum 64 leaves, learning rate 0.1, row subsampling 0.9, column subsampling 0.9, and seed 0. For point estimates of first-seen time, the raw baseline is the literal assumption that the COFF compile time is the file time; its error is therefore \(|\Delta_{\mathrm{days}}|\).

Each MalConv model was trained for one epoch with an 8 MB per-file input cap and batch size 32. Evaluation uses the MalConv2 low-memory convolution path: each batch is read once, the max-pooling winners are identified by chunked scans, and the selected receptive-window slices are recomputed for the final representation. 

Below, we will detail how we encode the loss and prediction for both the XGBoost and MalConv models. Multiple approaches are taken, treating the problem as a binary classification problem and an ordinal regression problem where we view the goal as predicting a discrete time range (e.g., true time is within 90 days of COFF timestamp), as a regression problem looking at directly regressing the date in absolute and relative terms, as a ranking problem where we try only to properly order samples by their first-seen date, and as a Bayesian model where we attempt to marginalize the partially observed set of good/bad timestamps.  

\subsection{Threshold Classification}
\label{sec:threshold-method}

The threshold classifiers predict whether the raw COFF timestamp is within a fixed tolerance of first seen. For threshold \(q\in\{\leq30,\leq60,\leq90,\leq180\}\) days, the binary target is
\[
  Y^{(q)}_i = \mathbb{1}\{|\Delta_{\mathrm{days},i}| \leq q\}.
\]
The XGBoost objective is binary logistic loss. If \(s_i=f_q(X_i)\) is the model score and \(\hat{p}_i=\{1+\exp(-s_i)\}^{-1}\), the minimized loss is
\[
  \mathcal{L}_{\mathrm{bin}}^{(q)}
  =
  -\sum_i
  \left[
    Y^{(q)}_i \log \hat{p}_i
    +
    \left(1-Y^{(q)}_i\right)\log\left(1-\hat{p}_i\right)
  \right].
\]
The resulting score is a probability estimate for the within-threshold event. The raw baseline for this stage is the empirical share of files satisfying the threshold, not a learned score. We do not use a Multinomial loss because of the ordering of the events, so an Ordinal approach is more appropriate --- as discussed in the next section. 

\subsection{Ordinal Absolute-Delta Prediction}
\label{sec:ordinal-method}

The ordinal target is the bin index of \(|\Delta_{\mathrm{days}}|\). The bins are [0--30), [30--60), [60--90), [90--180), then repeated doubling of the final range until the observed maximum is covered. The largest bin is [23,040--46,080) days. Reported quasi-MAE maps predicted bin probabilities back to days using the median training-set magnitude in each bin.

The corrected ordinal model is a proportional-odds ordered logit. If \(B_i\in\{1,\ldots,K\}\) is the bin and \(\eta_i=f_{\mathrm{ord}}(X_i)\) is a scalar model score, then ordered cutpoints \(\theta_1<\cdots<\theta_{K-1}\) define
\[
  \Pr(B_i \le k \mid X_i)
  =
  \sigma(\theta_k-\eta_i),
  \qquad
  \sigma(a)=\{1+\exp(-a)\}^{-1}.
\]
The bin probabilities are
\[
  p_{ik} =
  \begin{cases}
    \sigma(\theta_1-\eta_i), & k=1,\\
    \sigma(\theta_k-\eta_i)-\sigma(\theta_{k-1}-\eta_i), & 1<k<K,\\
    1-\sigma(\theta_{K-1}-\eta_i), & k=K.
  \end{cases}
\]
Both the XGBoost and MalConv2 ordinal versions minimize the negative log likelihood
\[
  \mathcal{L}_{\mathrm{ord}}
  =
  -\sum_i \log p_{iB_i}.
\]
For XGBoost, \(f_{\mathrm{ord}}\) is the boosted-tree scalar margin and the cutpoints are initialized from empirical training-bin frequencies and calibrated on the training margins. For MalConv2, \(f_{\mathrm{ord}}\) is the scalar raw-byte network output and the cutpoints are learned jointly with the network. The point estimate for absolute delta is \(\sum_{k=1}^{K}p_{ik}r_k\), where \(r_k\) is the training-set median representative for bin \(k\). While more complex ordinal models exist, prior work has found that there is little benefit when using a sufficiently powerful non-linear learner~\citep{Lu2022}.

\subsection{Signed Delta Prediction}
\label{sec:delta-method}

The signed-delta approach predicts the offset from the COFF timestamp to the first-seen date, and its target is \(d_i=\Delta_{\mathrm{days},i}\). A predicted delta can be added to the observed COFF timestamp to form an implied first-seen estimate. The XGBoost signed-delta regressor uses the quantile objective with \(\alpha=0.5\), corresponding to median pinball loss. With residual \(u_i=d_i-\hat{d}_i\), the loss is
\[
  \mathcal{L}_{\Delta,\mathrm{xgb}}
  =
  \sum_i \rho_{0.5}(u_i),
  \qquad
  \rho_{\alpha}(u)=u\left(\alpha-\mathbb{1}\{u<0\}\right).
\]
The MalConv2 signed-delta model uses the same target after robust scaling. If \(\tilde{d}_i\) is the scaled target, the smooth \(L_1\) loss is
\[
  \mathcal{L}_{\Delta,\mathrm{mc}}
  =
  \sum_i h(\tilde{d}_i-\hat{\tilde{d}}_i),
  \qquad
  h(r)=
  \begin{cases}
    0.5r^2, & |r|<1,\\
    |r|-0.5, & \text{otherwise}.
  \end{cases}
\]
The minor discrepancy in the loss function is due to using the standard loss functions available in each package, and since neither yielded meaningfully improved results, we do not ablate them further for computational/time considerations. 

\subsection{Absolute First-Seen Prediction}
\label{sec:absolute-method}

The absolute-time target attempts to directly predict the date that the malware sample was first seen, independent of how it relates relative to the COFF compilation metadata. Its target is \(t_i=\texttt{first\_submission\_date}/86400\), in days since the Unix epoch. The XGBoost absolute-time regressor uses the same median pinball loss as the signed-delta regressor,
\[
  \mathcal{L}_{T,\mathrm{xgb}}
  =
  \sum_i \rho_{0.5}(t_i-\hat{t}_i).
\]
The MalConv2 absolute-time model uses robustly scaled targets and smooth \(L_1\) loss,
\[
  \mathcal{L}_{T,\mathrm{mc}}
  =
  \sum_i h(\tilde{t}_i-\hat{\tilde{t}}_i),
\]
with \(h(\cdot)\) defined as in Section~\ref{sec:delta-method}.

\subsection{Learning-to-Rank and Ordering Scores}
\label{sec:ranking-method}

One of our desires is to be able to sort samples by their first-seen dates to create train/test splits that minimize information leakage. Knowing/predicting the date of a sample, as the prior sections do in various ways, allows one to construct this and make informed inferences about the dates (presuming you can have them with sufficient precision). An acceptable alternative would be knowing the ordering of samples in time, without knowing the actual timestamp date itself. This is a hindrance in studying factors in time itself, but it would still be beneficial if possible. 

This is known as a learning-to-rank problem, where we attempt to use the timestamp of first-seen as the rank ordering itself. By training a model that produces a score that correlates (in the Spearman sense) with the first-seen date, one can solve the problem. 
Scores are induced by the raw COFF timestamp, threshold classifiers, ordinal models, delta models, absolute-time models, the PyMC model, and learned rankers.

Three different learning-to-rank models were trained with XGBoost's pairwise ranking objective. XGBoost's learning-to-rank functionality requires prespecifying groups of files that have rankings within the group. Each query group was (arbitrarily) defined to 128 files\footnote{Compute cost can increase quadratically with group size, so making a singular group is not viable.}, and the relevance labels are the within-group ranks of \(\texttt{first\_submission\_date}\), with later first-seen times assigned larger ranks. The grouping schemes are random groups across the full training time range, groups stratified by label and PE type, and local groups sampled within 30-day first-seen windows. Each ranking view uses 779,648 training rows, approximately one-third of the PE/.NET training rows. In this way, we have groups that capture large/random gaps (via random sampling), groups that attempt to infer type-specific factors (via stratification), and groups that attempt to infer small-scale time differences (by selecting within the 30-day window). 

For a query group \(G\), both the learned XGBoost rankers and the MalConv2 ranking model optimize pairwise logistic ordering terms of the form
\[
  \mathcal{L}_{\mathrm{rank}}
  =
  \sum_{G}
  \sum_{\substack{i,j\in G\\ F_i>F_j}}
  \log\left(1+\exp\left(-(s_i-s_j)\right)\right),
\]
where \(s_i\) is the model score and later first-seen files should receive larger scores. The MalConv2 ranking model receives raw bytes and forms the same pairwise loss over files in each batch.

\subsection{Latent Timestamp Reliability Model}
\label{sec:graphical-method}

Bayesian modeling is one way to incorporate our knowledge about the underlying problem structure to make more accurate or reliable inferences of the unknowns, or despite the unknowns, in a system, which has proven effective~\citep{klein_bayesian_2026,klein_more_2024,Robberechts_2021,Arora_2011,ijcai2015p107,Arous_2021,NIPS2016_46072631}. We attempt to apply the same insight here. 
Another way to view the problem is as two populations of samples: those with reliable COFF timestamps and those with unreliable/modified timestamps. We can infer that some timestamps are reliable (small gap between COFF and first-seen) and others are untruthful (e.g., a timestamp set to the future), with ambiguous files in between. This creates a partially observed hidden variable of reliability, for which we could try to make a Bayesian model.

A Bayesian model was fit to the EMBER24 feature vectors to make the reliability assumption explicit. For file \(i\), let \(F_i\) be first seen, \(C_i\) be the observed COFF timestamp, and \(X_i\) be the timestamp-removed EMBER features. The latent true compile age is
\[
  A_i = F_i - T_i > 0,
\]
where \(T_i\) is the unobserved true compile time. A latent indicator \(Z_i\) denotes whether the observed COFF timestamp is truthful or modified. The high-precision bad-timestamp rule is one-sided: if the rule fires, \(Z_i\) is forced to the modified state; otherwise \(Z_i\) remains latent. We use PyMC3 to fit this model~\citep{pymc-devs_pymc_2026}

The fitted prototype uses Bayesian Additive Regression Trees (BART) to keep the non-linear power of a tree-based model~\citep{tan_bayesian_2019,chipman_bayesian_1998,chipman_bart_2010}. A BART prior models \(\log A_i\) from source type, label, and the timestamp-removed Rich-header feature block. Conditional on \(Z_i=1\), the observed compile offset \(C_i-F_i\) follows a Student-\(t\) distribution centered at \(-A_i\). Conditional on \(Z_i=0\), \(C_i-F_i\) follows a broad Student-\(t\) artifact distribution. The mixture over \(Z_i\) is marginalized rather than sampled discretely. The per-row likelihood is
\[
  p(O_i \mid X_i,B_i,\Theta)
  =
  w_i\,t_{\nu_{\mathrm{truth}}}\!\left(O_i;-A_i,\sigma_{\mathrm{truth}}\right)
  +
  (1-w_i)\,
  t_{\nu_{\mathrm{mod}}}\!\left(O_i;\mu_{\mathrm{mod}},\sigma_{\mathrm{mod}}\right),
\]
where \(O_i=C_i-F_i\), \(A_i=\exp(\operatorname{BART}(X_i))\), and \(w_i=0\) for obvious-bad rows and \(w_i=p\) otherwise. In inference, PyMC samples from the posterior proportional to the product of the likelihood and the stated priors, rather than minimizing a point-estimate loss. The generative story used by the code is summarized here in Algorithm~\ref{alg:timestamp-generative-story}.

\begin{algorithm}[!h]
\caption{Latent timestamp reliability model}
\label{alg:timestamp-generative-story}
\begin{algorithmic}[1]
\Require Per-file predictors \(X_i\), obvious-bad rule \(B_i\), first seen \(F_i\), and COFF timestamp \(C_i\)
\Ensure Posterior over latent ages \(A_i\), mixture parameters, and timestamp-reliability probabilities
\State \(p \sim \operatorname{Beta}(2,2)\) \Comment{truthful probability for non-obvious-bad rows}
\State \(\sigma_T \sim \operatorname{HalfNormal}(\cdot)\)
\State \(\nu_T \sim 2+\operatorname{Exponential}(\cdot)\)
\State \(\mu_M \sim \operatorname{Normal}(\cdot)\)
\State \(\sigma_M \sim \operatorname{HalfNormal}(\cdot)\)
\State \(\nu_M \sim 2+\operatorname{Exponential}(\cdot)\)
\For{all files \(i\)}
  \State \(O_i \gets C_i-F_i\) \Comment{observed compile offset}
  \State \(g_i \gets \operatorname{BART}(X_i)\)
  \State \(g_i \gets \operatorname{clip}(g_i,\log A_{\min},\log A_{\max})\)
  \State \(A_i \gets \exp(g_i)\)
  \State \(T_i \gets F_i-A_i\) \Comment{enforces \(T_i\leq F_i\)}
  \State \(P_{T,i} \gets t_{\nu_T}(\mu=-A_i,\sigma=\sigma_T)\)
  \State \(P_M \gets t_{\nu_M}(\mu=\mu_M,\sigma=\sigma_M)\)
  \If{\(B_i=1\)}
    \State \(w_i \gets 0\)
  \Else
    \State \(w_i \gets p\)
  \EndIf
  \State \(O_i \sim w_iP_{T,i}+(1-w_i)P_M\) \Comment{\(Z_i\) marginalized}
\EndFor
\end{algorithmic}
\end{algorithm}

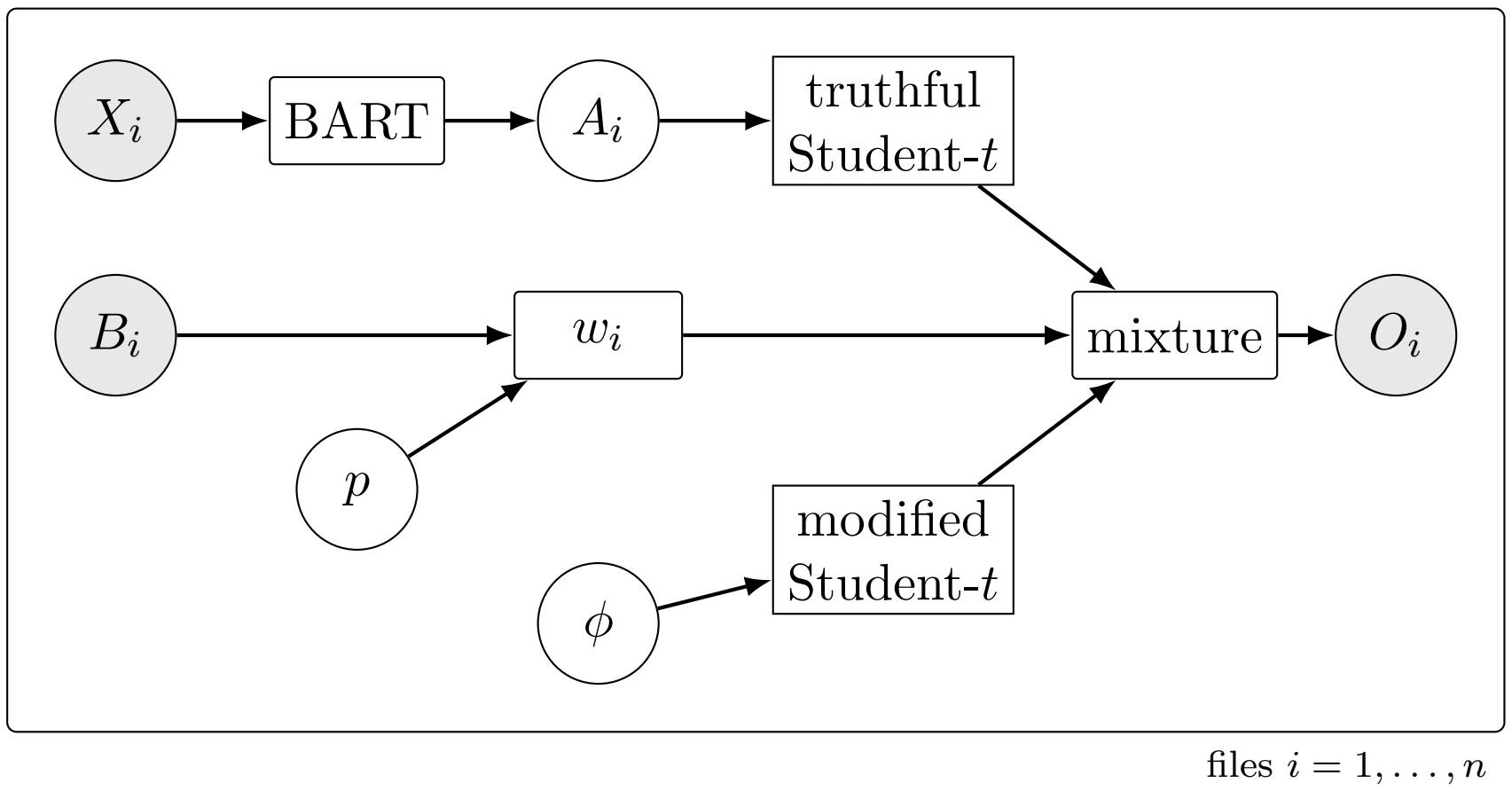
\begin{figure}[H]
\centering
\begin{tikzpicture}[
  x=1cm,
  y=1cm,
  latent/.style={circle, draw, minimum size=0.9cm, inner sep=1pt, align=center},
  obs/.style={circle, draw, fill=gray!18, minimum size=0.9cm, inner sep=1pt, align=center},
  det/.style={rectangle, draw, rounded corners=1pt, minimum width=1.25cm, minimum height=0.65cm, inner sep=3pt, align=center},
  dist/.style={rectangle, draw, minimum width=1.7cm, minimum height=0.65cm, inner sep=3pt, align=center},
  arr/.style={-{Latex[length=2mm]}, thick}
]
\node[obs] (x) at (0,1.6) {\(X_i\)};
\node[obs] (b) at (0,0) {\(B_i\)};
\node[latent] (p) at (1.8,-1.15) {\(p\)};

\node[det] (bart) at (1.8,1.6) {BART};
\node[latent] (a) at (3.6,1.6) {\(A_i\)};
\node[det] (w) at (3.6,0) {\(w_i\)};
\node[latent] (phi) at (3.6,-2.15) {\(\phi\)};

\node[dist] (truth) at (5.8,1.6) {truthful\\Student-\(t\)};
\node[dist] (mod) at (5.8,-1.6) {modified\\Student-\(t\)};
\node[det] (mix) at (7.9,0) {mixture};
\node[obs] (o) at (9.55,0) {\(O_i\)};

\draw[arr] (x) -- (bart);
\draw[arr] (bart) -- (a);
\draw[arr] (a) -- (truth);
\draw[arr] (phi) -- (mod);
\draw[arr] (b) -- (w);
\draw[arr] (p) -- (w);
\draw[arr] (w) -- (mix);
\draw[arr] (truth) -- (mix);
\draw[arr] (mod) -- (mix);
\draw[arr] (mix) -- (o);

\node[draw, rounded corners=2pt, fit=(x)(b)(p)(bart)(a)(w)(phi)(truth)(mod)(mix)(o), inner sep=10pt,
  label={[font=\scriptsize, anchor=north east]south east:files \(i=1,\ldots,n\)}] {};
\end{tikzpicture}
\caption{Graphical model for the timestamp reliability model in Algorithm~\ref{alg:timestamp-generative-story}. Distribution boxes denote the two Student-\(t\) likelihood components for \(O_i=C_i-F_i\); the mixture box denotes the marginalized likelihood that combines them using \(w_i\).}
\label{fig:graphical-model}
\end{figure}

The graphical structure corresponding to Algorithm~\ref{alg:timestamp-generative-story} is shown here in Figure~\ref{fig:graphical-model}.

The larger run used 2,000 random training rows, 50 BART trees, four chains, and 1,000 tune plus 1,000 posterior draws per chain. The saved trace has no divergences, maximum observed NUTS tree depth 5~\citep{JMLR:v15:hoffman14a}, and worst scalar \(\hat{R}=1.0083\), indicating the model has converged~\citep{gelman_inference_1992}.

\section{Results}
\label{sec:results}

Our initial modeling results show limited predictive utility. Below, we analyze each method on both the out-of-time test set and the EMBER2024 challenge set, finding consistently that accurate prediction of first-seen dates is difficult. These results motivate the reframed obvious-bad timestamp analysis in \autoref{sec:obviously-bad-shap}, which examines what can be learned about clearly unreliable timestamps.

\subsection{Macro Distribution of Timestamp Disagreement}
\label{sec:macro-results}

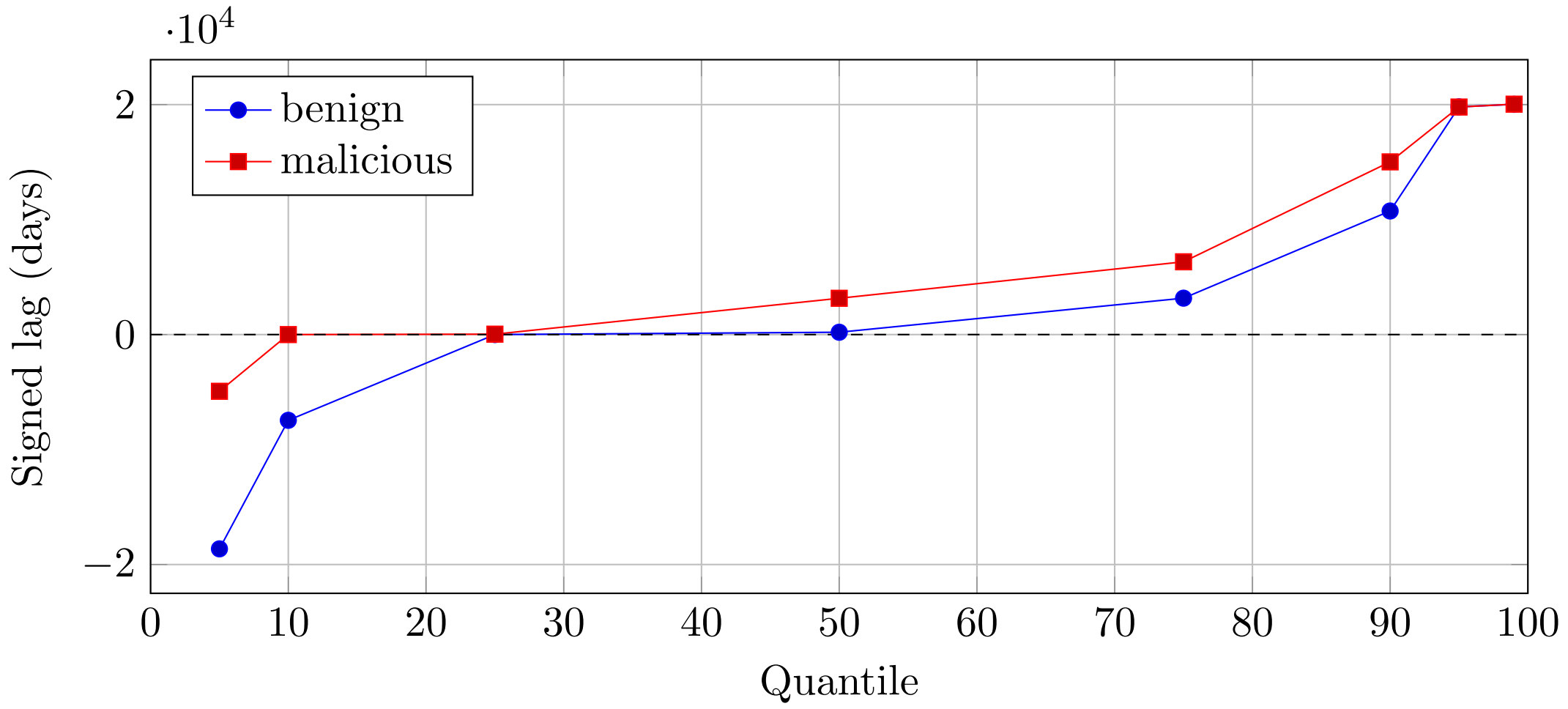
\begin{figure}[H]
\centering
\begin{tikzpicture}
\begin{axis}[
  width=0.92\linewidth,
  height=0.42\linewidth,
  xlabel={Quantile},
  ylabel={Signed lag (days)},
  xmin=0,
  xmax=100,
  grid=both,
  legend pos=north west,
  legend cell align=left,
]
\addplot+[mark=*] table[x=quantile,y=days] {tables/lag_quantiles_benign.dat};
\addlegendentry{benign}
\addplot+[mark=square*] table[x=quantile,y=days] {tables/lag_quantiles_malicious.dat};
\addlegendentry{malicious}
\addplot[dashed, black] coordinates {(0,0) (100,0)};
\end{axis}
\end{tikzpicture}
\caption{Signed lag quantiles for benign and malicious PE/.NET files. Positive values indicate that the compile timestamp predates first seen. Negative values indicate a compile timestamp after first seen.}
\label{fig:lag-quantiles}
\end{figure}

The signed-lag distribution differs by label. The median signed lag is 203.5 days for benign files and 3,157.3 days for malicious files. After excluding negative lags (i.e., file claims to be time traveling from the future), the median lag is 516.0 days for benign files and 3,646.8 days for malicious files. The upper tail is large for both labels: the 95th percentile signed lag is 19,804.6 days for benign files and 19,789.9 days for malicious files.
The signed-lag quantiles are shown here in Figure~\ref{fig:lag-quantiles}.

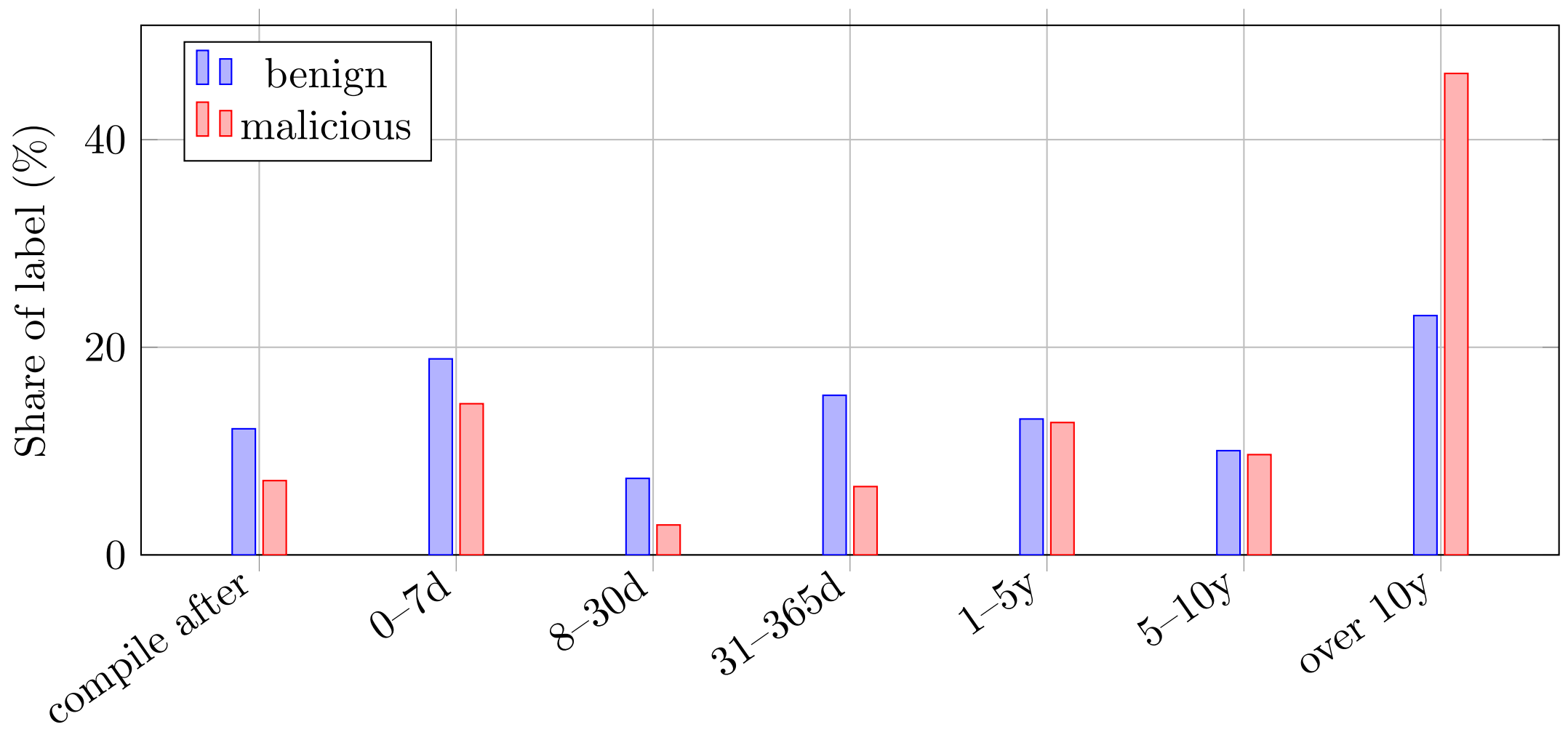
\begin{figure}[H]
\centering
\begin{tikzpicture}
\begin{axis}[
  ybar,
  width=0.95\linewidth,
  height=0.42\linewidth,
  ylabel={Share of label (\%)},
  xtick={0,1,2,3,4,5,6},
  xticklabels={compile after,0--7d,8--30d,31--365d,1--5y,5--10y,over 10y},
  x tick label style={rotate=35, anchor=east},
  ymin=0,
  bar width=6pt,
  grid=major,
  legend pos=north west,
]
\addplot table[x expr=\coordindex,y=share] {tables/lag_category_shares_benign.dat};
\addlegendentry{benign}
\addplot table[x expr=\coordindex,y=share] {tables/lag_category_shares_malicious.dat};
\addlegendentry{malicious}
\end{axis}
\end{tikzpicture}
\caption{Macro lag-category shares by label. The ``compile after'' category counts rows where the COFF timestamp is after first seen.}
\label{fig:lag-categories}
\end{figure}

The rate of time-traveling samples that claim to be compiled after the file was seen is non-trivial in both benign and malicious cases, which surprised us. A discretization of the lag-category shares is shown here in Figure~\ref{fig:lag-categories}. These results indicate that, largely, both benign and malicious samples have roughly comparable rates of all kinds of erroneous timestamps, even if the malicious files have a propensity for greater magnitude lags. 

\subsection{Threshold Classification}
\label{sec:threshold-results}

The classifier results from \autoref{sec:threshold-method} are shown here in Table~\ref{tab:test-classifiers} for the regular test split and Table~\ref{tab:challenge-classifiers} for the post-train challenge subset. The raw-share column is the fraction of rows in which the raw COFF timestamp falls within the threshold. 

\begin{table}[H]
\centering
\caption{Regular test split threshold classification. With-COFF models received the observed COFF timestamp as an input feature; no-COFF models did not.}
\label{tab:test-classifiers}
\scriptsize
\begin{tabular}{lrrrrr}
\toprule
Threshold & Raw share & No-COFF AUC & No-COFF F1 & With-COFF AUC & With-COFF F1 \\
\midrule
30d & 0.2211 & 0.9705 & 0.8337 & 0.9756 & 0.8596 \\
60d & 0.2459 & 0.9762 & 0.8606 & 0.9734 & 0.8863 \\
90d & 0.2623 & 0.9774 & 0.8690 & 0.9743 & 0.9039 \\
180d & 0.2967 & 0.9810 & 0.8947 & 0.9849 & 0.9419 \\
\bottomrule
\end{tabular}
\end{table}

\begin{table}[H]
\centering
\caption{Post-train challenge threshold classification. Raw share is the fraction of post-train challenge rows whose COFF timestamp is within the threshold.}
\label{tab:challenge-classifiers}
\scriptsize
\begin{tabular}{lrrrrr}
\toprule
Threshold & Raw share & No-COFF AUC & No-COFF F1 & With-COFF AUC & With-COFF F1 \\
\midrule
30d & 0.3504 & 0.9734 & 0.8085 & 0.9794 & 0.9378 \\
60d & 0.4132 & 0.9671 & 0.8270 & 0.9844 & 0.9448 \\
90d & 0.4579 & 0.9680 & 0.8634 & 0.9595 & 0.9450 \\
180d & 0.5587 & 0.9810 & 0.8967 & 0.9832 & 0.9796 \\
\bottomrule
\end{tabular}
\end{table}

In both cases, we can see that having the COFF information in the feature vector is beneficial to the model (e.g., one can easily flag nonsensically forward/backward-looking timestamps), and that the ability to predict within a finite thresholded range is possible with high accuracy. However, most samples have larger timestamps, so the ability to find a subset of reliable dates does not allow us to sort the samples that fall outside the small-threshold classification model. When factoring in this component of evaluation, as we do in the next section, we see performance degrades significantly. 

\subsection{Point Prediction by Target}
\label{sec:point-results}

In this section, we now evaluate our ability to meaningfully predict when a sample was first seen, which requires more resolution than a simple ``COFF timestamp is correct within $x$ days''. We had four modeling approaches to this: ordinal prediction of the time range, predicting the signed delta from the COFF timestamp, direct prediction of the absolute first-seen date, and our latent Bayesian model. 

Regular test point-estimation errors are shown here in Table~\ref{tab:test-point}. Rows are grouped by target or loss family. Raw COFF is the baseline first-seen estimate, so its point error is \(|\Delta_{\mathrm{days}}|\). The mean absolute error (MAE), median absolute error (MedAE), and 90th-percentile absolute error are shown for each case. The simplest result is that our Bayesian model degrades to using the COFF timestamp directly, and a Bayesian approach will require more significant work to be meaningful. 

\begin{table}[H]
\centering
\caption{Regular test split point-estimation errors, in days, grouped by target or loss family.}
\label{tab:test-point}
\scriptsize
\resizebox{\linewidth}{!}{%
\begin{tabular}{lllrrr}
\toprule
Target/loss & Representation & COFF as predictor & MAE & MedAE & p90 \\
\midrule
raw timestamp & COFF baseline & -- & 6,396.68 & 2,681.86 & 20,026.00 \\
\addlinespace
ordinal abs-delta & XGBoost & no & 2,747.66 & 1,486.89 & 7,163.55 \\
ordinal abs-delta & XGBoost & yes & 2,044.67 & 1,147.65 & 7,593.38 \\
ordinal abs-delta & MalConv2 & raw bytes & 3,294.74 & 1,548.88 & 9,886.17 \\
\addlinespace
signed delta & XGBoost & no & 2,029.49 & 274.33 & 5,351.69 \\
signed delta & XGBoost & yes & 296.33 & 154.05 & 781.46 \\
signed delta & MalConv2 & raw bytes & 2,474.56 & 465.46 & 8,196.86 \\
\addlinespace
absolute first-seen & XGBoost & no & 164.65 & 190.61 & 262.43 \\
absolute first-seen & XGBoost & yes & 148.71 & 162.79 & 266.71 \\
absolute first-seen & MalConv2 & raw bytes & 202.65 & 215.63 & 274.45 \\
\addlinespace
latent age score & PyMC-BART & no & 6,394.95 & 2,678.77 & 20,024.39 \\
\bottomrule
\end{tabular}
}
\end{table}

The large improvement of the with-COFF signed-delta model is expected because the signed-delta target is defined directly using the COFF timestamp. We interpret this result as a sanity check demonstrating the impact of restoring direct access to a target-defining field, rather than as evidence of an independent source of temporal information. More broadly, the other approaches can improve upon the raw COFF timestamp baseline, which is encouraging. However, their errors remain on the order of at least half a year. Because the EMBER2024 corpus was collected within a single calendar year, errors of this magnitude substantially limit our ability to organize samples when first-seen information is unavailable. Indeed, predicting approximately the median collection date would produce an error on a similar scale because samples are distributed roughly across the year. These results indicate that significant additional work, potentially including new feature engineering, is needed to address this problem effectively. For example, external information about source code or library evolution could provide landmarks for identifying what code ``looks like'' at different points in time~\citep{liu_assemblage_2024}.

Post-train challenge set errors are shown here in Table~\ref{tab:challenge-point}. The table is restricted to 605 challenge rows whose first-seen dates are later than the maximum first-seen date in the training split. The results are consistent with the standard test set, though, surprisingly, the absolute first-seen error is lower for the challenge files than for the standard test set. 

\begin{table}[H]
\centering
\caption{Post-train challenge point-estimation errors, in days, grouped by target or loss family.}
\label{tab:challenge-point}
\scriptsize
\resizebox{\linewidth}{!}{%
\begin{tabular}{lllrrr}
\toprule
Target/loss & Representation & COFF as predictor & MAE & MedAE & p90 \\
\midrule
raw timestamp & COFF baseline & -- & 2,591.26 & 127.01 & 10,817.47 \\
\addlinespace
ordinal abs-delta & XGBoost & no & 3,008.55 & 3,103.88 & 4,685.11 \\
ordinal abs-delta & XGBoost & yes & 1,453.19 & 1,119.91 & 2,169.12 \\
ordinal abs-delta & MalConv2 & raw bytes & 6,953.36 & 1,451.44 & 20,000.93 \\
\addlinespace
signed delta & XGBoost & no & 1,997.83 & 1,056.79 & 5,150.41 \\
signed delta & XGBoost & yes & 311.82 & 181.83 & 716.57 \\
signed delta & MalConv2 & raw bytes & 7,186.90 & 1,776.78 & 20,026.84 \\
\addlinespace
absolute first-seen & XGBoost & no & 97.39 & 60.44 & 215.93 \\
absolute first-seen & XGBoost & yes & 70.34 & 47.72 & 187.56 \\
absolute first-seen & MalConv2 & raw bytes & 232.34 & 244.10 & 294.10 \\
\addlinespace
latent age score & PyMC-BART & no & 2,589.90 & 124.37 & 10,819.26 \\
\bottomrule
\end{tabular}
}
\end{table}

\subsection{Sorting Ability by Score Type}
\label{sec:time-ordering-results}

Although the error is large relative to the timescale over which the data is collected, this does not necessarily mean the models would be ineffective at sorting the data by first-seen date. If the models have low variance but high bias, they could still be effective sorters. We test this hypothesis in this section, along with the direct learning-to-rank model in \autoref{sec:ranking-method}.

Table~\ref{tab:time-ordering} summarizes all the results. Rows in each held-out split were ordered by the score and compared with the true first-seen order. Pair accuracy is the probability that two sampled rows are ordered the same way by the score and by true first seen. Spearman is the rank correlation between prediction and ground truth. Rank error is the absolute difference between a row's predicted-order position and true first-seen position, normalized by split size. The challenge rows in Table~\ref{tab:time-ordering} are restricted to the post-train challenge subset. Ultimately, all models in all cases are ineffective at producing a useful ordering of the data, with a maximum Spearman correlation of 0.22. 

\begin{table}[H]
\centering
\caption{Temporal ordering induced by model scores, grouped by score family. Challenge rows are restricted to files first seen after the train maximum.}
\label{tab:time-ordering}
\scriptsize
\resizebox{\linewidth}{!}{%
\begin{tabular}{lllrrrr}
\toprule
Split & Score family & Representation & Pair acc. & Spearman & Median rank err. & Median position-time err. \\
\midrule
test & raw timestamp & COFF baseline & 0.5346 & 0.1041 & 0.2822 & 23.73 \\
test & latent age score & PyMC-BART & 0.5334 & 0.0852 & 0.2789 & 23.44 \\
test & within90 probability & XGBoost no-COFF & 0.5287 & 0.0838 & 0.2591 & 21.83 \\
test & within90 probability & XGBoost with-COFF & 0.5208 & 0.0603 & 0.2739 & 23.10 \\
test & ordinal abs-delta & XGBoost no-COFF & 0.5109 & 0.0309 & 0.2832 & 23.88 \\
test & ordinal abs-delta & XGBoost with-COFF & 0.5103 & 0.0298 & 0.2803 & 23.63 \\
test & ordinal abs-delta & MalConv2 & 0.5308 & 0.0945 & 0.2793 & 23.58 \\
test & signed delta & XGBoost no-COFF & 0.5024 & 0.0064 & 0.3027 & 25.56 \\
test & signed delta & XGBoost with-COFF & 0.5093 & 0.0227 & 0.3012 & 25.37 \\
test & signed delta & MalConv2 & 0.4941 & -0.0185 & 0.3069 & 25.89 \\
test & absolute first-seen & XGBoost no-COFF & 0.4990 & -0.0051 & 0.3052 & 25.77 \\
test & absolute first-seen & XGBoost with-COFF & 0.4974 & -0.0080 & 0.3064 & 25.85 \\
test & absolute first-seen & MalConv2 & 0.4514 & -0.1461 & 0.3113 & 26.27 \\
test & ranking & XGBoost local no-COFF & 0.5019 & 0.0062 & 0.2941 & 24.80 \\
test & ranking & XGBoost local with-COFF & 0.5014 & 0.0034 & 0.2951 & 24.88 \\
test & ranking & MalConv2 & 0.4561 & -0.1326 & 0.3092 & 26.10 \\
\addlinespace
challenge post-train & raw timestamp & COFF baseline & 0.5835 & 0.2168 & 0.2566 & 19.67 \\
challenge post-train & latent age score & PyMC-BART & 0.5840 & 0.2203 & 0.2566 & 19.57 \\
challenge post-train & within90 probability & XGBoost no-COFF & 0.5201 & 0.0661 & 0.2947 & 22.91 \\
challenge post-train & within90 probability & XGBoost with-COFF & 0.4818 & -0.0498 & 0.2980 & 22.62 \\
challenge post-train & ordinal abs-delta & XGBoost no-COFF & 0.4703 & -0.0873 & 0.3245 & 24.82 \\
challenge post-train & ordinal abs-delta & XGBoost with-COFF & 0.4925 & -0.0227 & 0.3046 & 22.90 \\
challenge post-train & ordinal abs-delta & MalConv2 & 0.5221 & 0.0597 & 0.2881 & 21.80 \\
challenge post-train & signed delta & XGBoost no-COFF & 0.4428 & -0.1735 & 0.3560 & 27.13 \\
challenge post-train & signed delta & XGBoost with-COFF & 0.5123 & 0.0379 & 0.2864 & 21.83 \\
challenge post-train & signed delta & MalConv2 & 0.4989 & -0.0029 & 0.2732 & 21.40 \\
challenge post-train & absolute first-seen & XGBoost no-COFF & 0.4694 & -0.0903 & 0.3344 & 25.39 \\
challenge post-train & absolute first-seen & XGBoost with-COFF & 0.4827 & -0.0505 & 0.3079 & 23.55 \\
challenge post-train & absolute first-seen & MalConv2 & 0.4820 & -0.0548 & 0.3146 & 24.30 \\
challenge post-train & ranking & XGBoost local no-COFF & 0.5108 & 0.0323 & 0.3046 & 23.15 \\
challenge post-train & ranking & XGBoost local with-COFF & 0.4903 & -0.0294 & 0.3013 & 22.40 \\
challenge post-train & ranking & MalConv2 & 0.4824 & -0.0508 & 0.3262 & 24.77 \\
\bottomrule
\end{tabular}
}
\end{table}

\begin{figure}[H]
\centering
\begin{tikzpicture}
\begin{groupplot}[
  group style={group size=2 by 1, horizontal sep=1.2cm},
  width=0.48\linewidth,
  height=0.40\linewidth,
  ybar,
  xtick={0,1,2,3,4,5,6},
  xticklabels={COFF,PyMC,absolute,delta,ordinal,rank-local,within90},
  x tick label style={rotate=35, anchor=east},
  grid=major,
  legend pos=north west,
]
\nextgroupplot[
  ymin=0.45,
  ymax=0.75,
  ylabel={Pair ordering accuracy},
  bar width=4pt,
]
\addplot table[x expr=\coordindex,y=test_pair] {tables/time_ordering_metrics.dat};
\addlegendentry{test}
\addplot table[x expr=\coordindex,y=challenge_pair] {tables/time_ordering_metrics.dat};
\addlegendentry{challenge post-train}

\nextgroupplot[
  ymin=0,
  ymax=0.34,
  ylabel={Median rank error},
  bar width=4pt,
]
\addplot table[x expr=\coordindex,y=test_rankerr] {tables/time_ordering_metrics.dat};
\addplot table[x expr=\coordindex,y=challenge_rankerr] {tables/time_ordering_metrics.dat};
\end{groupplot}
\end{tikzpicture}
\caption{Ordering quality from model scores on the regular test split and post-train challenge subset. Lower rank error is better.}
\label{fig:time-ordering-bars}
\end{figure}
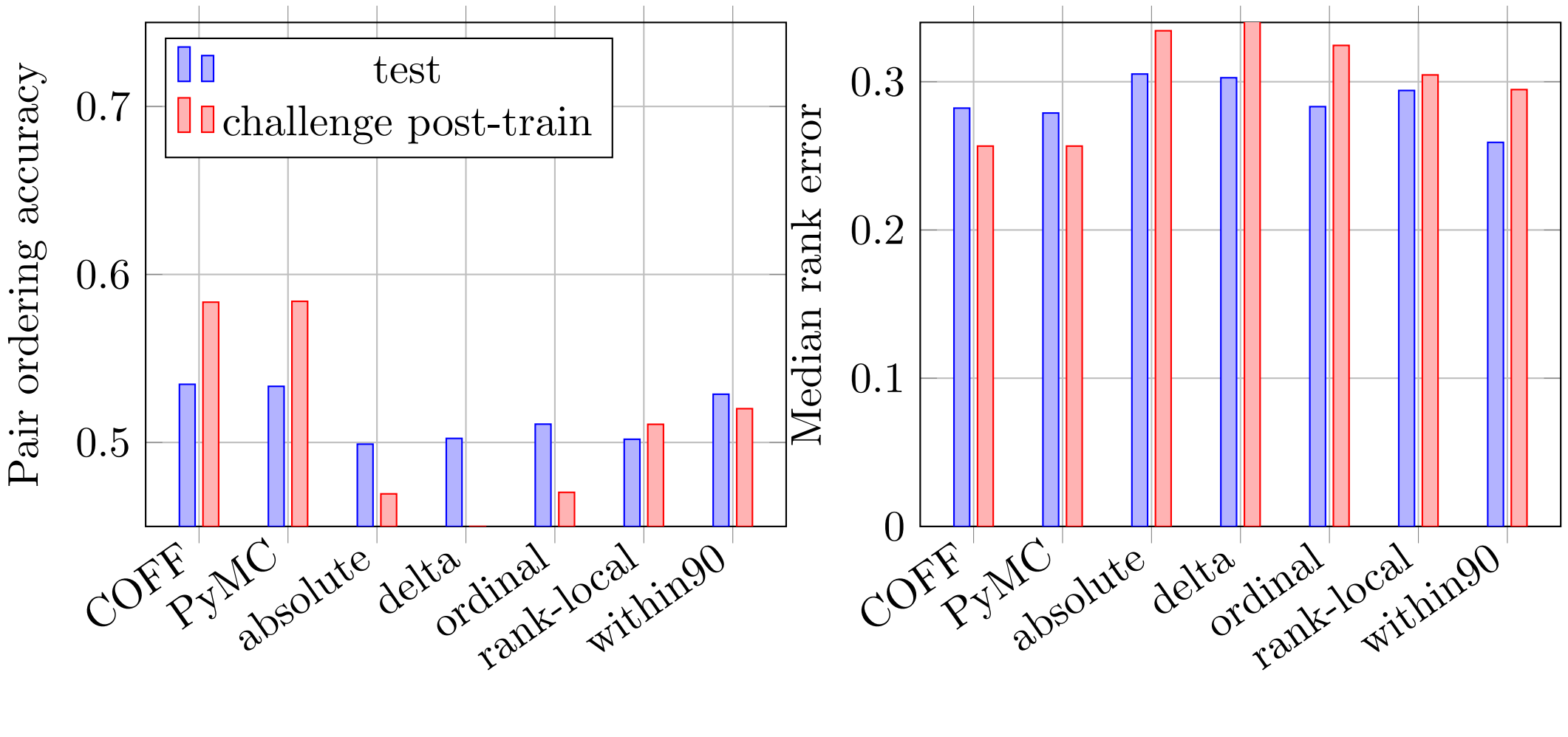

The same ordering metrics are summarized graphically here in Figure~\ref{fig:time-ordering-bars}. This visualization makes it more obvious that there is no particular bias in favor/against any singular method between test/challenge samples: each approach is equally bad. 

\begin{figure}[H]
\centering
\begin{tikzpicture}
\begin{groupplot}[
  group style={group size=2 by 1, horizontal sep=1.2cm},
  width=0.48\linewidth,
  height=0.40\linewidth,
  xlabel={Predicted-order percentile},
  ylabel={Median true first-seen offset (days)},
  xmin=0,
  xmax=100,
  grid=both,
  legend pos=north west,
  legend cell align=left,
]
\nextgroupplot[
  title={regular test},
  ymin=0,
  ymax=90,
]
\addplot+[black, dashed, mark=none] table[x=pct,y=oracle_offset_p50_days] {tables/time_ordering_curve_test_oracle.dat};
\addlegendentry{oracle}
\addplot+[mark=*] table[x=pct,y=true_offset_p50_days] {tables/time_ordering_curve_test_compile.dat};
\addlegendentry{COFF}
\addplot+[mark=diamond*] table[x=pct,y=true_offset_p50_days] {tables/time_ordering_curve_test_pymc.dat};
\addlegendentry{PyMC}
\addplot+[mark=square*] table[x=pct,y=true_offset_p50_days] {tables/time_ordering_curve_test_absolute.dat};
\addlegendentry{absolute}
\addplot+[mark=triangle*] table[x=pct,y=true_offset_p50_days] {tables/time_ordering_curve_test_ranklocal.dat};
\addlegendentry{rank-local}
\nextgroupplot[
  title={challenge post-train},
  ymin=0,
  ymax=90,
]
\addplot+[black, dashed, mark=none] table[x=pct,y=oracle_offset_p50_days] {tables/time_ordering_curve_challenge_oracle.dat};
\addplot+[mark=*] table[x=pct,y=true_offset_p50_days] {tables/time_ordering_curve_challenge_compile.dat};
\addplot+[mark=diamond*] table[x=pct,y=true_offset_p50_days] {tables/time_ordering_curve_challenge_pymc.dat};
\addplot+[mark=square*] table[x=pct,y=true_offset_p50_days] {tables/time_ordering_curve_challenge_absolute.dat};
\addplot+[mark=triangle*] table[x=pct,y=true_offset_p50_days] {tables/time_ordering_curve_challenge_ranklocal.dat};
\end{groupplot}
\end{tikzpicture}
\caption{Chronology curves after sorting rows by selected model scores. Each point is the median true first-seen offset within a predicted-order bin. The oracle line is the corresponding true sorted order. The challenge panel is restricted to the post-train subset.}
\label{fig:time-ordering-curves}
\end{figure}
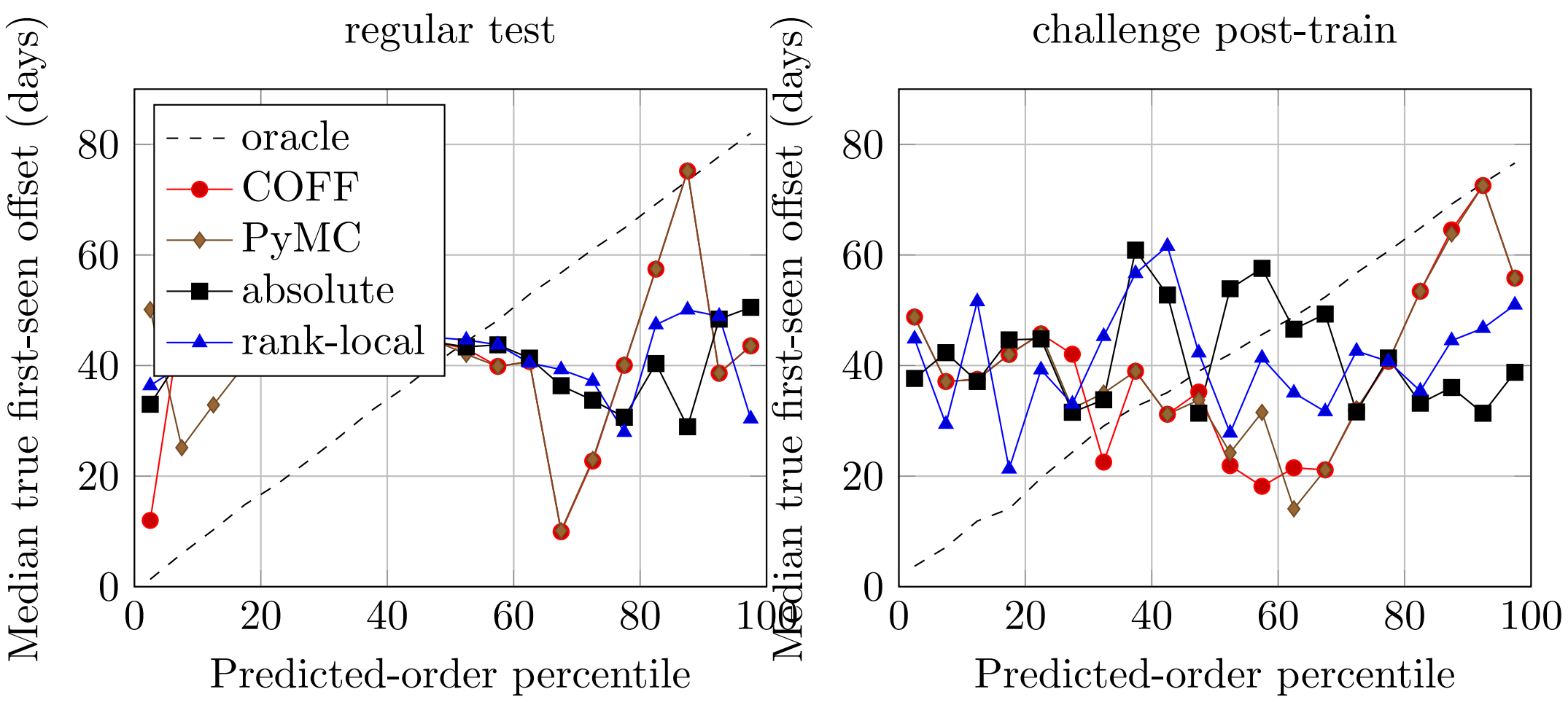

One of my hypotheses is that the models are more effective at ranking one span of time than others due to an issue or transient factor, which would provide a foothold for more meaningful exploration and sample reversal to determine how to improve this factor. The predicted-order chronology curves are shown here in Figure~\ref{fig:time-ordering-curves}, which shows that all models are generally errant uniformly over time.

\subsubsection{Latent Timestamp Reliability}
\label{sec:pymc-results}

The graphical model in \figureref{fig:graphical-model} separates the sampled rows into two regimes. COFF timestamps close to first seen have high posterior probability of being truthful. Multi-year lags, future timestamps, and rows matched by the obvious-bad rule are assigned to the modified/artifact component. In the 2,000-row posterior analysis, 671 rows were marked obviously bad by rule and therefore had zero posterior truthful probability by construction. Among the remaining 1,329 rows, the posterior mean truthful probability was 0.3975, but the distribution was bimodal: the median was 0.0060 and the 75th percentile was 0.9978.

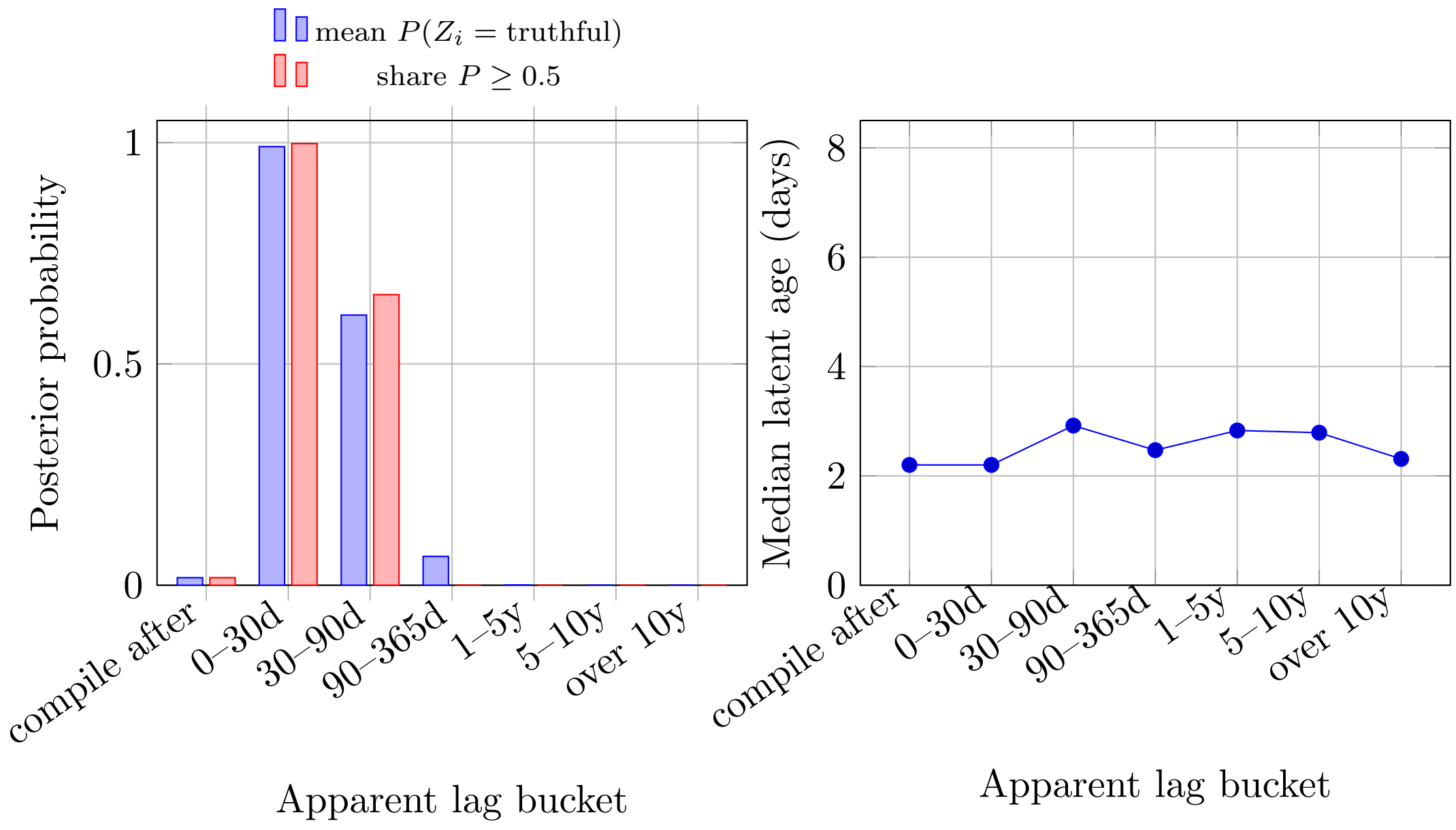
\begin{figure}[H]
\centering
\begin{tikzpicture}
\begin{groupplot}[
  group style={group size=2 by 1, horizontal sep=1.1cm},
  width=0.48\linewidth,
  height=0.40\linewidth,
  xtick=data,
  x tick label style={rotate=35, anchor=east},
  grid=major,
]
\nextgroupplot[
  ybar,
  ymin=0,
  ymax=1.05,
  ylabel={Posterior probability},
  xlabel={Apparent lag bucket},
  xticklabels={compile after,0--30d,30--90d,90--365d,1--5y,5--10y,over 10y},
  bar width=7pt,
  legend style={
    at={(0.5,1.03)},
    anchor=south,
    font=\scriptsize,
    draw=none
  },
]
\addplot table[x expr=\coordindex,y=mean_truth] {tables/pymc_truth_by_delta.dat};
\addlegendentry{mean \(P(Z_i=\mathrm{truthful})\)}
\addplot table[x expr=\coordindex,y=share_ge_0_5] {tables/pymc_truth_by_delta.dat};
\addlegendentry{share \(P\geq 0.5\)}
\nextgroupplot[
  ymin=0,
  ymax=8.5,
  ylabel={Median latent age (days)},
  xlabel={Apparent lag bucket},
  xticklabels={compile after,0--30d,30--90d,90--365d,1--5y,5--10y,over 10y},
]
\addplot+[mark=*] table[x expr=\coordindex,y=median_age_days] {tables/pymc_truth_by_delta.dat};
\end{groupplot}
\end{tikzpicture}
\caption{Posterior interpretation of the graphical timestamp model by apparent lag bucket. The left panel shows posterior truthful-timestamp probability and the share of rows with probability at least 0.5. The right panel shows the posterior median latent compile age before first seen.}
\label{fig:pymc-truth-by-delta}
\end{figure}

The posterior reliability summaries by apparent lag bucket are shown here in Figure~\ref{fig:pymc-truth-by-delta}.

The corresponding posterior group summaries are shown here in Table~\ref{tab:pymc-group-results}.

\begin{table}[H]
\centering
\caption{Interpretable posterior summaries from the graphical timestamp model. ``Bad rule'' rows are forced into the modified/artifact component.}
\label{tab:pymc-group-results}
\scriptsize
\begin{tabular}{lrrrrr}
\toprule
Group & \(N\) & Mean \(P_{\mathrm{truth}}\) & Median \(P_{\mathrm{truth}}\) & Share \(P_{\mathrm{truth}}\geq 0.5\) & Median \(A_i\) \\
 & & & & & days \\
\midrule
all & 2,000 & 0.2641 & 0.0000 & 0.2640 & 2.39 \\
not obviously bad & 1,329 & 0.3975 & 0.0060 & 0.3973 & 2.47 \\
bad rule & 671 & 0.0000 & 0.0000 & 0.0000 & 2.20 \\
benign & 1,033 & 0.3154 & 0.0002 & 0.3175 & 2.78 \\
malicious & 967 & 0.2093 & 0.0000 & 0.2068 & 2.03 \\
DotNet & 209 & 0.3730 & 0.0001 & 0.3780 & 2.20 \\
Win32 & 1,325 & 0.1857 & 0.0000 & 0.1864 & 2.73 \\
Win64 & 466 & 0.4382 & 0.0530 & 0.4335 & 1.50 \\
\bottomrule
\end{tabular}
\end{table}

The scalar mixture parameters are consistent with this separation. The posterior mean truthful timestamp noise is 4.39 days. The modified/artifact component has posterior mean location \(C_i-F_i=-3{,}288\) days and scale 4,242 days. By apparent lag bucket, the posterior mean truthful probability is 0.9910 for 0--30 day lags, 0.6104 for 30--90 day lags, 0.0651 for 90--365 day lags, and effectively zero for lags beyond one year. The latent age estimates should therefore be interpreted conditionally on the model's artifact assumption: the posterior median \(A_i\) is 2.39 days, which reflects assigning large disagreements to modification rather than to very old true compile times.

\subsubsection{Absolute Time Prediction and Extrapolation}

\begin{figure}[H]
\centering
\begin{tikzpicture}
\begin{groupplot}[
  group style={group size=2 by 1, horizontal sep=1.8cm},
  width=0.48\linewidth,
  height=0.40\linewidth,
  ymin=0,
  ymax=340,
  grid=both,
  legend style={at={(0.03,0.03)}, anchor=south west, legend columns=2, font=\scriptsize, fill=white, fill opacity=0.85, draw opacity=1, text opacity=1},
  legend cell align=left,
]
\nextgroupplot[
  title={regular test},
  xtick={0,1,2,3},
  xticklabels={Sep,Oct,Nov,Dec},
  xlabel={first-seen month, 2024},
  ylabel={Absolute error (days)},
]
\addplot+[mark=*] table[x expr=\coordindex,y=p50] {tables/absolute_error_by_month_test.dat};
\addlegendentry{p50}
\addplot+[mark=square*] table[x expr=\coordindex,y=p75] {tables/absolute_error_by_month_test.dat};
\addlegendentry{p75}
\addplot+[mark=triangle*] table[x expr=\coordindex,y=p90] {tables/absolute_error_by_month_test.dat};
\addlegendentry{p90}
\addplot+[mark=diamond*] table[x expr=\coordindex,y=p95] {tables/absolute_error_by_month_test.dat};
\addlegendentry{p95}
\addplot+[black, densely dashed, mark=none, forget plot] coordinates {(0,0) (0,340)};
\node[anchor=south west, font=\scriptsize] at (axis cs:0,305) {train max};
\nextgroupplot[
  title={challenge},
  xtick={0,2,4,6,8,10,12,14},
  xticklabels={Sep,Nov,Jan,Mar,May,Jul,Sep,Nov},
  xlabel={first-seen month},
  x tick label style={rotate=35, anchor=east},
]
\addplot+[mark=*] table[x expr=\coordindex,y=p50] {tables/absolute_error_by_month_challenge.dat};
\addplot+[mark=square*] table[x expr=\coordindex,y=p75] {tables/absolute_error_by_month_challenge.dat};
\addplot+[mark=triangle*] table[x expr=\coordindex,y=p90] {tables/absolute_error_by_month_challenge.dat};
\addplot+[mark=diamond*] table[x expr=\coordindex,y=p95] {tables/absolute_error_by_month_challenge.dat};
\addplot+[black, densely dashed, mark=none, forget plot] coordinates {(12,0) (12,340)};
\node[anchor=south west, font=\scriptsize] at (axis cs:12,305) {train max};
\end{groupplot}
\end{tikzpicture}
\caption{Absolute first-seen prediction absolute-error quantiles by true first-seen month. The dashed line marks the month containing the training maximum first-seen date, 2024-09-22.}
\label{fig:absolute-error-time}
\end{figure}
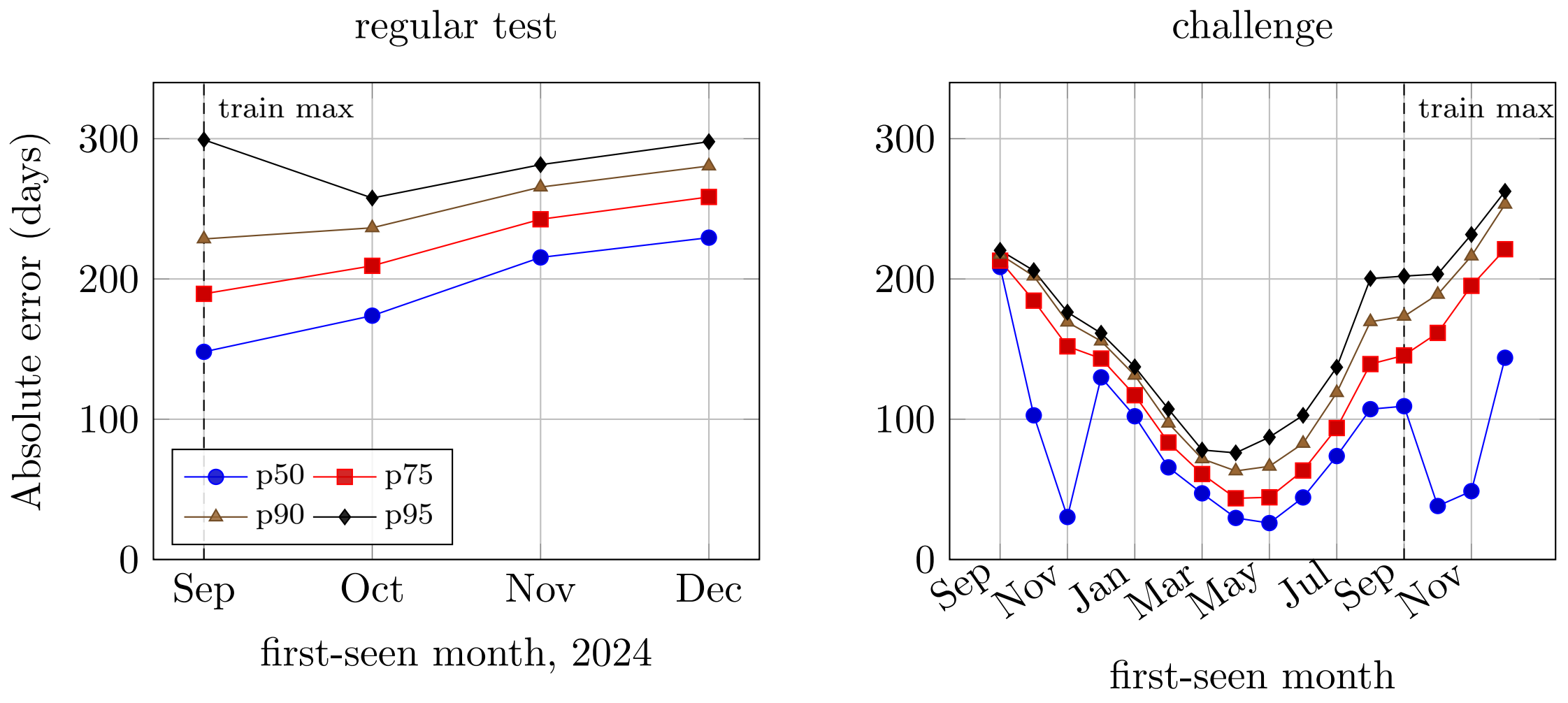

The absolute first-seen regressor has lower aggregate error than the delta-days regressor, but this does not imply reliable extrapolation. The training first-seen target range is 2023-09-24 to 2024-09-22. In the regular test split, 539,863 of 539,869 rows are after the training maximum first-seen date, but only 51,061 predictions are after that maximum. The model usually predicts into the training period and underpredicts later first-seen times.
We can see this by looking at the absolute-time error curves by first-seen month, which are shown here in Figure~\ref{fig:absolute-error-time}. As one might expect, the error across all thresholds increases over time (barring some sample-size favoritism toward the median error). 

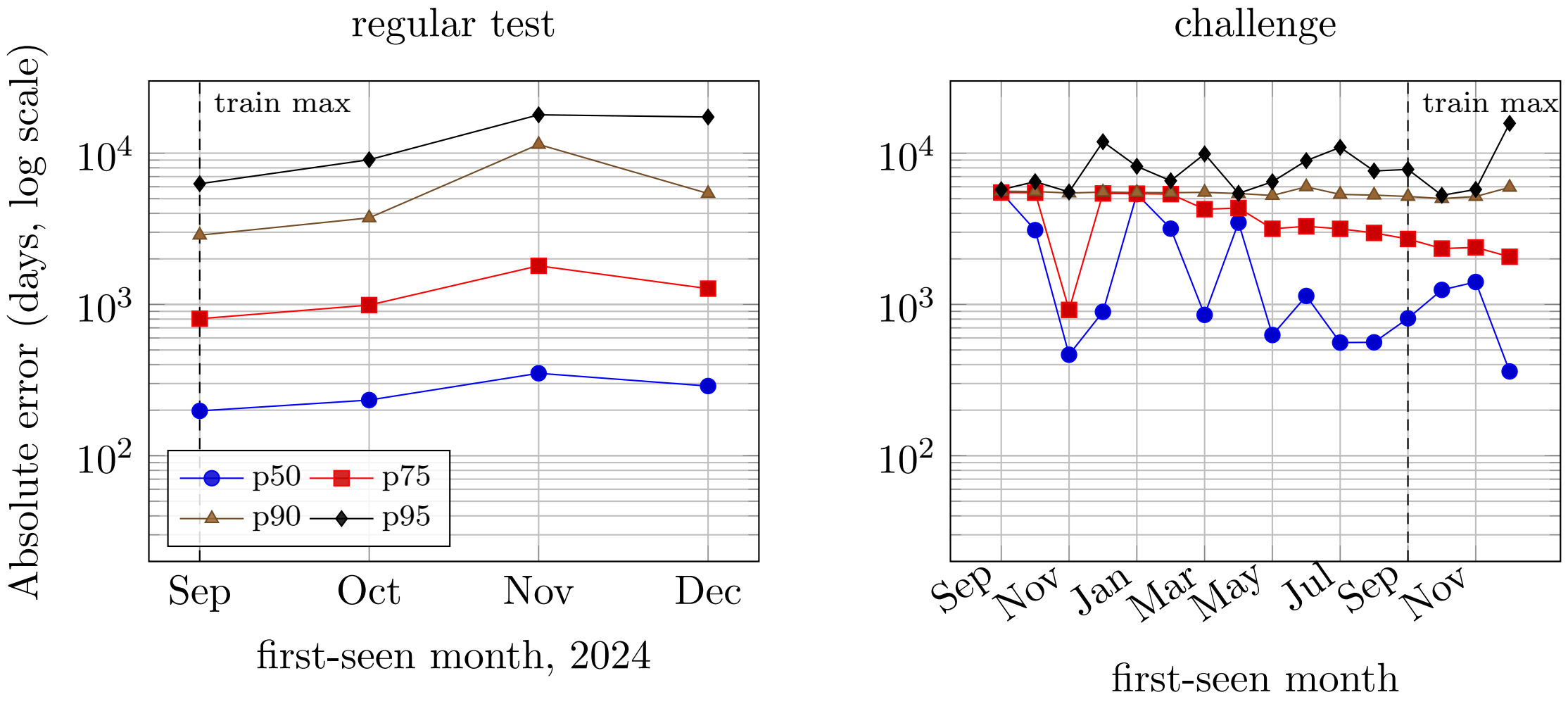
\begin{figure}[H]
\centering
\begin{tikzpicture}
\begin{groupplot}[
  group style={group size=2 by 1, horizontal sep=1.8cm},
  width=0.48\linewidth,
  height=0.40\linewidth,
  ymode=log,
  ymin=20,
  ymax=30000,
  grid=both,
  legend style={at={(0.03,0.03)}, anchor=south west, legend columns=2, font=\scriptsize, fill=white, fill opacity=0.85, draw opacity=1, text opacity=1},
  legend cell align=left,
]
\nextgroupplot[
  title={regular test},
  xtick={0,1,2,3},
  xticklabels={Sep,Oct,Nov,Dec},
  xlabel={first-seen month, 2024},
  ylabel={Absolute error (days, log scale)},
]
\addplot+[mark=*] table[x expr=\coordindex,y=p50] {tables/delta_error_by_month_test.dat};
\addlegendentry{p50}
\addplot+[mark=square*] table[x expr=\coordindex,y=p75] {tables/delta_error_by_month_test.dat};
\addlegendentry{p75}
\addplot+[mark=triangle*] table[x expr=\coordindex,y=p90] {tables/delta_error_by_month_test.dat};
\addlegendentry{p90}
\addplot+[mark=diamond*] table[x expr=\coordindex,y=p95] {tables/delta_error_by_month_test.dat};
\addlegendentry{p95}
\addplot+[black, densely dashed, mark=none, forget plot] coordinates {(0,20) (0,30000)};
\node[anchor=south west, font=\scriptsize] at (axis cs:0,15000) {train max};
\nextgroupplot[
  title={challenge},
  xtick={0,2,4,6,8,10,12,14},
  xticklabels={Sep,Nov,Jan,Mar,May,Jul,Sep,Nov},
  xlabel={first-seen month},
  x tick label style={rotate=35, anchor=east},
]
\addplot+[mark=*] table[x expr=\coordindex,y=p50] {tables/delta_error_by_month_challenge.dat};
\addplot+[mark=square*] table[x expr=\coordindex,y=p75] {tables/delta_error_by_month_challenge.dat};
\addplot+[mark=triangle*] table[x expr=\coordindex,y=p90] {tables/delta_error_by_month_challenge.dat};
\addplot+[mark=diamond*] table[x expr=\coordindex,y=p95] {tables/delta_error_by_month_challenge.dat};
\addplot+[black, densely dashed, mark=none, forget plot] coordinates {(12,20) (12,30000)};
\node[anchor=south west, font=\scriptsize] at (axis cs:12,15000) {train max};
\end{groupplot}
\end{tikzpicture}
\caption{Delta-days prediction absolute-error quantiles by true first-seen month. The dashed line marks the month containing the training maximum first-seen date, 2024-09-22.}
\label{fig:delta-error-time}
\end{figure}

The corresponding delta-error curves are shown here in Figure~\ref{fig:delta-error-time}. In this case, the error is so large that it is more consistent over time, reflecting a lack of signal/efficacy in the model rather than any robustness to shifts in time. 

Rows where both the first-seen date and compile timestamp are after the training maximum provide the strongest evidence that the file time is outside the training target window. There are 178,671 such rows in the regular test split and 266 in the challenge split. The highest-margin examples have first-seen and compile dates in December 2024 that agree within roughly one day, while the absolute first-seen model often predicts months earlier. This is consistent with the monthly error curves in \figureref{fig:absolute-error-time}.

\section{Obvious-Bad SHAP Analysis}
\label{sec:obviously-bad-shap}

The preceding results indicate that the available features contain some information about sample age: models can predict whether COFF and first-seen timestamps fall within specified tolerances of 30--180 days, even though producing a useful general chronological ordering remains difficult. Motivated by this distinction, we reframe the problem as predicting whether a COFF timestamp is ``obviously bad'' according to a high-precision rule. This formulation yields a stronger predictive result and supports a more detailed analysis of the associated feature signals.

We will train one XGBoost classifier for the high-precision obvious-bad timestamp rule used in the Bayesian experiments. The positive label of ``obvious-bad'' is when the COFF timestamp is zero, more than one day after first seen, before 1995-01-01 UTC, on or after 2030-01-01 UTC, or appears at least 1,000 times as an exact timestamp value in the PE/.NET timestamp corpus\footnote{The PE COFF \texttt{TimeDateStamp} field has one-second nominal resolution, so exact duplicate values at this multiplicity are unlikely. This may need adjustment for different-sized corpora, and we did not ablate this due to runtime.}. The model uses the timestamp-removed 2,567-dimensional PE/.NET feature matrix and is trained on the benign and malicious train rows. We will then look at the error rates between benign/malicious samples, and SHapley Additive exPlanations (SHAP)~\citep{NIPS2017_7062,strumbelj_general_2011,Strumbelj:2014:EPM:2687513.2687579} explanations to better understand how benign and malicious samples differ in being ``obviously-bad''.

\subsection{Obvious-Bad Type Rates}
\label{sec:obviously-bad-type-rates}

The obvious-bad rule applies to both benign and malicious files in the train and regular-test splits. The overall rule rates are shown here in Table~\ref{tab:obvious-bad-overall-rates}. The displayed type columns omit the on-or-after-2030 condition because, for these train/test rows, it is redundant with the compile-after-first-seen condition.

\begin{table}[H]
\centering
\caption{Overall obvious-bad timestamp rates by split and label.}
\label{tab:obvious-bad-overall-rates}
\scriptsize
\begin{tabular}{llrrr}
\toprule
Split & Label & \(N\) & Obvious-bad \(N\) & Rate \\
\midrule
train & benign & 1,170,000 & 282,924 & 0.2418 \\
train & malicious & 1,170,000 & 531,602 & 0.4544 \\
test & benign & 270,000 & 88,033 & 0.3260 \\
test & malicious & 270,000 & 129,198 & 0.4785 \\
\bottomrule
\end{tabular}
\end{table}

The non-exclusive type rates among obvious-bad rows are shown here in Table~\ref{tab:obvious-bad-type-rates}. Negative cases are excluded from this table. A row can satisfy more than one obvious-bad condition, so rows can sum to more than one.

\begin{table}[H]
\centering
\caption{Non-exclusive obvious-bad type rates among PE/.NET rows marked obvious-bad in each split-label group. The repeated column counts exact COFF timestamps observed at least 1,000 times in the compact PE/.NET timestamp corpus.}
\label{tab:obvious-bad-type-rates}
\scriptsize
\begin{tabular}{llrrrr}
\toprule
Split & Label & Zero & After first seen & Before 1995 & Repeated \\
\midrule
train & benign & 0.2122 & 0.4865 & 0.3499 & 0.4277 \\
train & malicious & 0.1189 & 0.1419 & 0.3827 & 0.8604 \\
test & benign & 0.4226 & 0.4140 & 0.5131 & 0.5149 \\
test & malicious & 0.2246 & 0.1559 & 0.4693 & 0.8753 \\
\bottomrule
\end{tabular}
\end{table}

The obvious-bad classifier performance is shown here in Table~\ref{tab:obviously-bad-performance}. The challenge split contains only malicious rows, so the benign--malicious SHAP comparison below uses the regular test split. Here we can see high AUCs with F1 scores better than the non-COFF threshold results from \autoref{sec:threshold-results}, indicating a potential improvement in problem specification with respect to what we have the information and data to model, though some calibration failure given the F1 score is relatively low compared to AUROC. 

\begin{table}[H]
\centering
\caption{Obvious-bad timestamp classifier performance. The model is a single XGBoost binary classifier trained on benign and malicious train rows.}
\label{tab:obviously-bad-performance}
\scriptsize
\begin{tabular}{llrrrrrr}
\toprule
Split & Group & \(N\) & Bad rate & AUROC & AUPRC & Brier & F1@0.5 \\
\midrule
test & all & 539,869 & 0.4024 & 0.9943 & 0.9921 & 0.0318 & 0.9341 \\
test & benign & 269,873 & 0.3262 & 0.9958 & 0.9929 & 0.0176 & 0.9672 \\
test & malicious & 269,996 & 0.4785 & 0.9923 & 0.9916 & 0.0459 & 0.9098 \\
challenge & malicious & 4,868 & 0.3640 & 0.9970 & 0.9946 & 0.0356 & 0.9348 \\
\bottomrule
\end{tabular}
\end{table}

We further analyze this binary XGBoost model using SHAP, specifically TreeSHAP~\citep{lundberg_local_2020,yang_fast_2022}. TreeSHAP decomposes the model margin, not the final probability. If \(f(x_i)\) is the model's log-odds score for file \(i\), then
\[
  f(x_i) = \phi_0 + \sum_{j=1}^{p}\phi_{ij},
  \qquad
  \hat{p}_i = \frac{1}{1+\exp\{-f(x_i)\}},
\]
where \(\phi_0\) is the expected model margin and \(\phi_{ij}\) is the SHAP contribution of feature \(j\). A positive \(\phi_{ij}\) increases the odds assigned to the obvious-bad class by a factor \(\exp(\phi_{ij})\) conditional on the other contributions in the additive decomposition; a negative value decreases those odds. 

Our corpus has over 2000 features, which is too numerous for meaningful individual inspection. Thus, we will consider the natural groups of features by type/nature (e.g., entropy features are a single group). For a feature group \(G\), the signed group contribution is
$\Phi_{iG}=\sum_{j\in G}\phi_{ij}$.

The benign--malicious comparison reported below is
  $\Delta_G =
  \mathbb{E}[\Phi_{iG}\mid \mathrm{malicious}]
  -
  \mathbb{E}[\Phi_{iG}\mid \mathrm{benign}]$,
estimated on the full regular-test split. Thus, \(\Delta_G>0\) means that the same model assigns a larger obvious-bad log-odds contribution to malicious rows through feature group \(G\); \(\Delta_G<0\) means the contribution is larger for benign rows. These quantities describe the fitted predictor's additive evidence, not a causal effect of the feature group.

\begin{figure}[H]
\centering
\begin{tikzpicture}
\begin{axis}[
  ybar,
  width=0.95\linewidth,
  height=0.46\linewidth,
  ylabel={Malicious minus benign mean signed SHAP},
  xtick=data,
  xticklabels={header,section,imports,rich,strings,authenticode,warnings,histogram,data dirs,byte entropy,general,exports},
  x tick label style={rotate=35, anchor=east},
  grid=major,
  bar width=6pt,
]
\addplot table[x expr=\coordindex,y=signed_delta] {tables/obviously_bad_family_shap_test.dat};
\addplot+[black, dashed, mark=none, forget plot] coordinates {(-0.5,0) (11.5,0)};
\end{axis}
\end{tikzpicture}
\caption{Feature-family TreeSHAP differences for the obvious-bad timestamp classifier on the regular test split. Positive values indicate larger mean signed contribution for malicious rows; negative values indicate larger mean signed contribution for benign rows.}
\label{fig:obviously-bad-family-shap}
\end{figure}
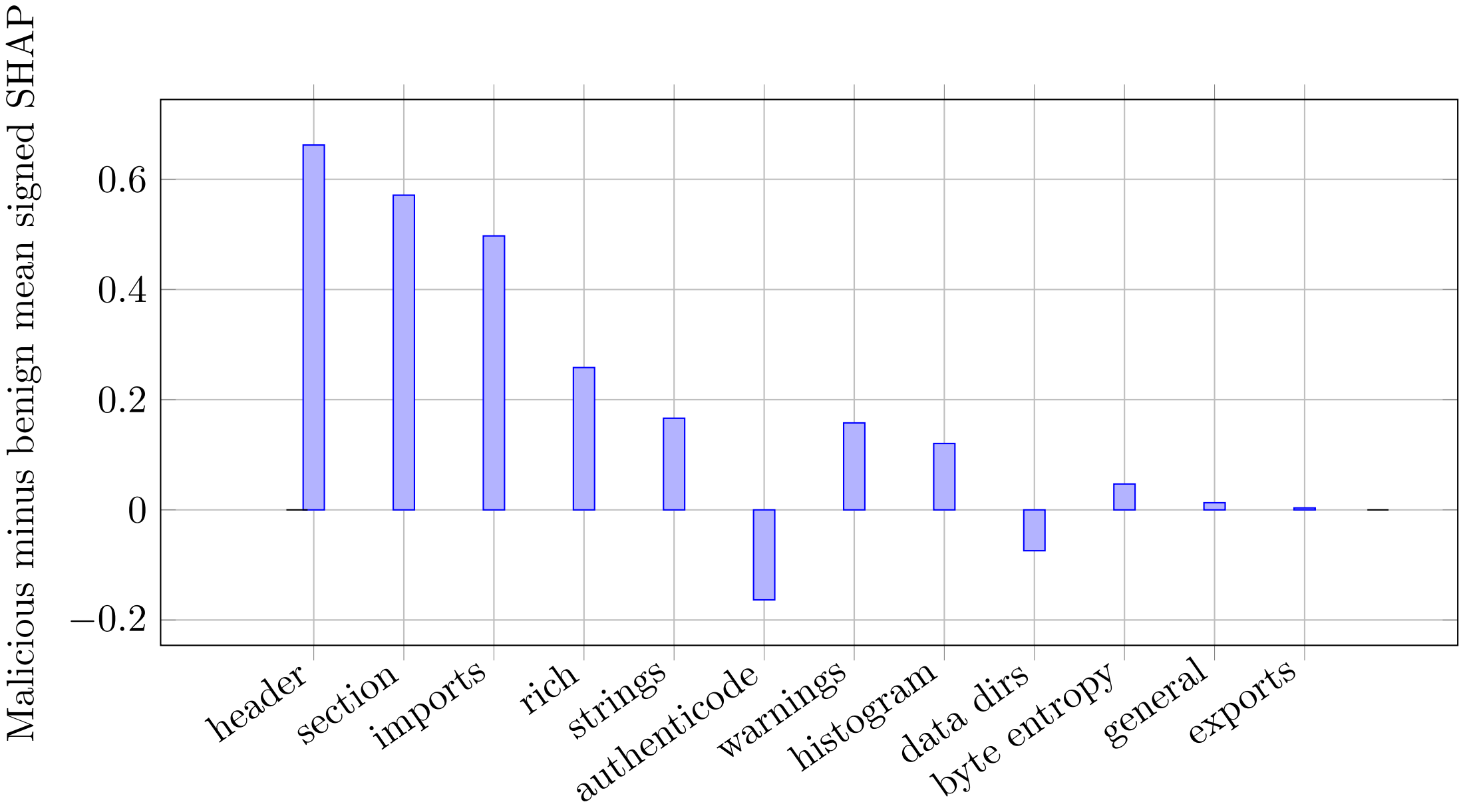

TreeSHAP values were computed for all 539,869 regular-test rows and all 4,868 challenge rows. Feature-family contributions sum the signed SHAP values within each EMBER feature block. Figure~\ref{fig:obviously-bad-family-shap} shows the malicious-minus-benign difference in mean signed family contribution under the model. It is immediately clear that some sections are meaningfully biased in which kinds of features they contribute predictive efficacy to. 

The corresponding feature-family values are shown here in Table~\ref{tab:obviously-bad-family-shap}. Signed values are in log-odds contribution units for the XGBoost binary classifier.

\begin{table}[H]
\centering
\caption{Regular-test feature-family TreeSHAP summaries for the obvious-bad timestamp classifier. \(\Delta\) is malicious minus benign mean signed SHAP.}
\label{tab:obviously-bad-family-shap}
\scriptsize
\begin{tabular}{lrrr}
\toprule
Feature family & Benign signed & Malicious signed & \(\Delta\) \\
\midrule
header & -0.6775 & -0.0152 & 0.6623 \\
section & -0.4576 & 0.1136 & 0.5712 \\
imports & -0.3140 & 0.1834 & 0.4973 \\
Rich header & -0.3393 & -0.0810 & 0.2583 \\
strings & -0.1188 & 0.0476 & 0.1664 \\
authenticode & 0.0649 & -0.0986 & -0.1634 \\
PE warnings & -0.0808 & 0.0771 & 0.1578 \\
histogram & -0.1158 & 0.0046 & 0.1204 \\
data directories & 0.0347 & -0.0395 & -0.0742 \\
byte entropy & -0.0628 & -0.0158 & 0.0470 \\
general & -0.0144 & -0.0015 & 0.0129 \\
exports & -0.0030 & 0.0005 & 0.0035 \\
\bottomrule
\end{tabular}
\end{table}

\begin{figure}[H]
\centering
\begin{tikzpicture}
\begin{groupplot}[
  group style={group size=2 by 1, horizontal sep=0.9cm},
  height=0.38\linewidth,
  ymin=-0.5,
  ymax=0.3,
  grid=both,
  clip=false,
  tick label style={font=\scriptsize},
]
\nextgroupplot[
  width=0.36\linewidth,
  height=0.38\linewidth,
  ylabel={Malicious minus benign mean signed SHAP},
  xmin=1,
  xmax=400,
  xtick={1,100,200,300,400},
]
\addplot+[mark=none, restrict x to domain=1:400] table[x=rank,y=malicious_minus_benign_signed] {tables/obviously_bad_all_features_signedrank_test.dat};
\addplot+[black, dashed, mark=none, forget plot] coordinates {(1,0) (400,0)};
\draw[thick] (rel axis cs:1,0.03) -- (rel axis cs:1.03,-0.03);
\draw[thick] (rel axis cs:1,0.10) -- (rel axis cs:1.03,0.04);
\nextgroupplot[
  width=0.36\linewidth,
  height=0.38\linewidth,
  xmin=2200,
  xmax=2567,
  xtick={2200,2400,2567},
  yticklabels={},
]
\addplot+[mark=none, restrict x to domain=2200:2567] table[x=rank,y=malicious_minus_benign_signed] {tables/obviously_bad_all_features_signedrank_test.dat};
\addplot+[black, dashed, mark=none, forget plot] coordinates {(2200,0) (2567,0)};
\draw[thick] (rel axis cs:-0.03,0.03) -- (rel axis cs:0.03,-0.03);
\draw[thick] (rel axis cs:-0.03,0.10) -- (rel axis cs:0.03,0.04);
\end{groupplot}
\node[anchor=north] at ($(group c1r1.south east)!0.5!(group c2r1.south west)+(0,-0.45cm)$) {Feature rank by signed benign--malicious difference};
\end{tikzpicture}
\caption{Regular-test feature-level SHAP differences for the obvious-bad timestamp classifier, sorted by malicious-minus-benign mean signed SHAP, with ranks 401--2199 omitted from the x-axis display. Values above zero indicate features with larger obvious-bad log-odds contribution for malicious rows; values below zero indicate larger contribution for benign rows.}
\label{fig:obviously-bad-all-feature-shap}
\end{figure}
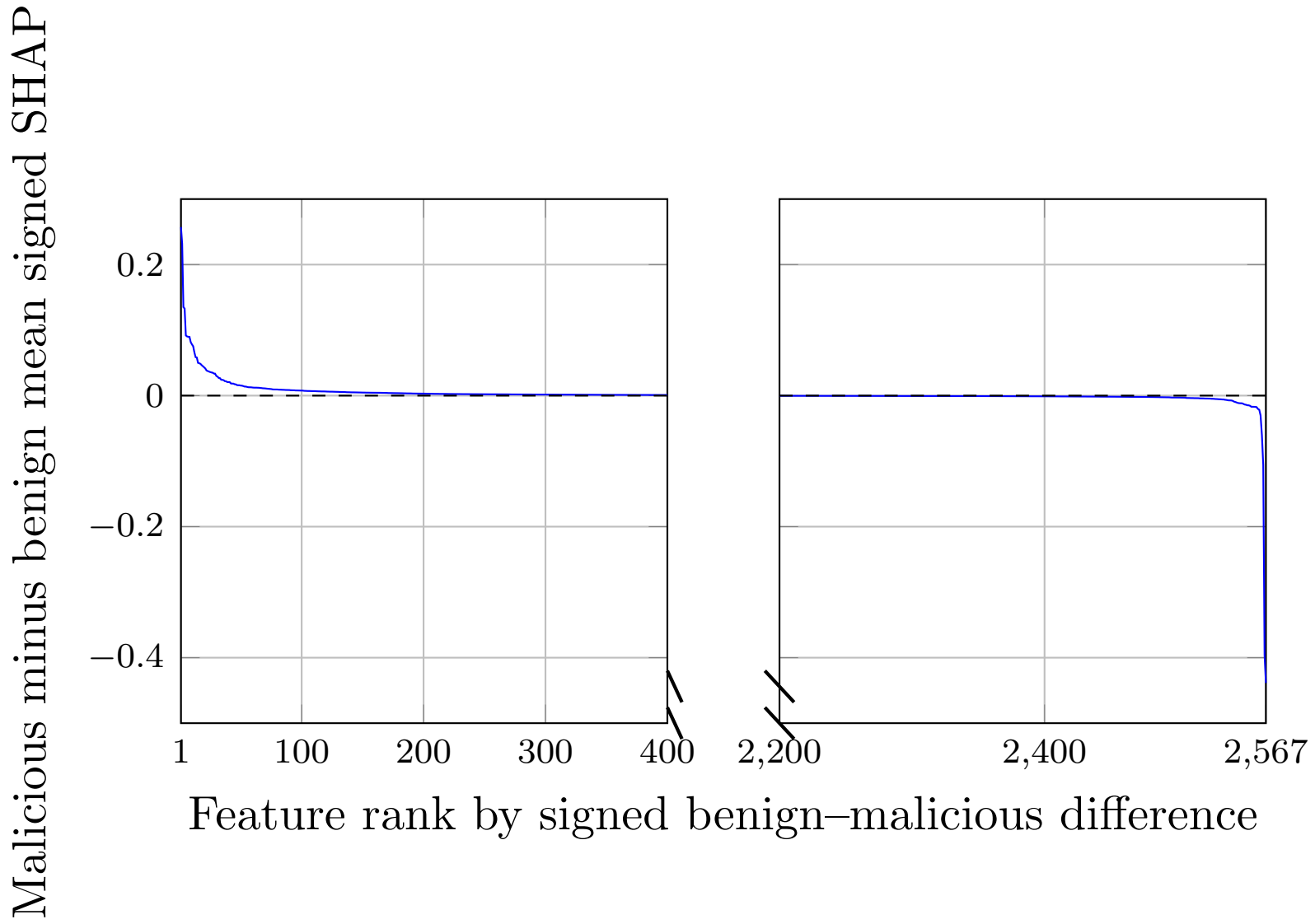

The feature-level signed differences for all 2,567 timestamp-removed features are shown here in Figure~\ref{fig:obviously-bad-all-feature-shap}. The features are sorted from the largest difference between malicious and benign contributions to the smallest. This indicates that the vast majority of features are non-informative to the problem at hand, but a small minority drives all predictive utility -- indicating a small subset of each group is worth looking at. 

The largest magnitude individual signed-SHAP differences are shown here in Table~\ref{tab:obviously-bad-top-features}. Several offsets map directly to named PE fields. The import and section entries are hashed bins produced by the EMBER-style feature extractor, so those rows identify the hashed subvector rather than a single reversible API or section name. Reverse engineering how these different features are impacted by various methods/processes/side effects of how timestamps are set is beyond the capacity of our group to consider at this time.

\begin{table}[H]
\centering
\caption{Largest regular-test individual-feature signed TreeSHAP differences for the obvious-bad timestamp classifier.}
\label{tab:obviously-bad-top-features}
\scriptsize
\begin{tabular}{rllrrr}
\toprule
Rank & Feature & Column & Benign & Malicious & \(\Delta\) \\
\midrule
1 & Authenticode signing-time minus COFF time & 2478 & 0.2658 & -0.1728 & -0.4386 \\
2 & Debug directory size & 2416 & 0.3000 & -0.0955 & -0.3956 \\
3 & Optional header uninitialized-data size & 714 & -0.0907 & 0.1665 & 0.2572 \\
4 & Authenticode latest signing time & 2477 & -0.1541 & 0.0778 & 0.2319 \\
5 & PE warning: suspicious section flags & 2546 & -0.0641 & 0.0710 & 0.1351 \\
6 & Optional header major linker version & 704 & -0.2105 & -0.0776 & 0.1329 \\
7 & Optional header major image version & 702 & 0.0481 & -0.0576 & -0.1057 \\
8 & Import function hash bin 1225 & 2218 & -0.0402 & 0.0516 & 0.0918 \\
9 & Export directory virtual address & 2405 & -0.0468 & 0.0436 & 0.0903 \\
10 & Import function hash bin 976 & 1969 & -0.0211 & 0.0687 & 0.0898 \\
11 & COFF characteristic: executable image & 726 & -0.0004 & 0.0893 & 0.0897 \\
12 & Rich-header pair count & 2438 & -0.0675 & 0.0141 & 0.0816 \\
13 & Import function hash bin 670 & 1663 & -0.0348 & 0.0434 & 0.0782 \\
14 & Export directory size & 2404 & -0.0550 & 0.0204 & 0.0754 \\
15 & Load-config directory virtual address & 2425 & -0.0274 & 0.0387 & 0.0661 \\
\bottomrule
\end{tabular}
\end{table}

In particular, we suspect that these relationships are not causal -- there is nothing that must be altered in an executable, or its headers, to alter the timestamp. Thus, the fact that there are strongly predictive features for this task is an indicator that they are correlative in quality, and artifacts of processes/tools leaking information about the original actors or build environment~\citep{zak_ngrams_2017}. We suspect that such an investigation is non-trivial.

\subsection{Association With Original Malware Scores}
\label{sec:original-model-time-artifact-association}

An alternative hypothesis may be that our model is learning something correlated in time with the general benign/malicious prediction problem, rather than artifacts about the timestamp itself. We test this by looking for a correlation between the original LightGBM models' predictions of benign/malicious and the SHAP scores we have extracted. If the SHAP scores and benign/malicious scores are correlated, then it is an indicator that we have inadvertently learned a benign/malicious model (but with extra steps).

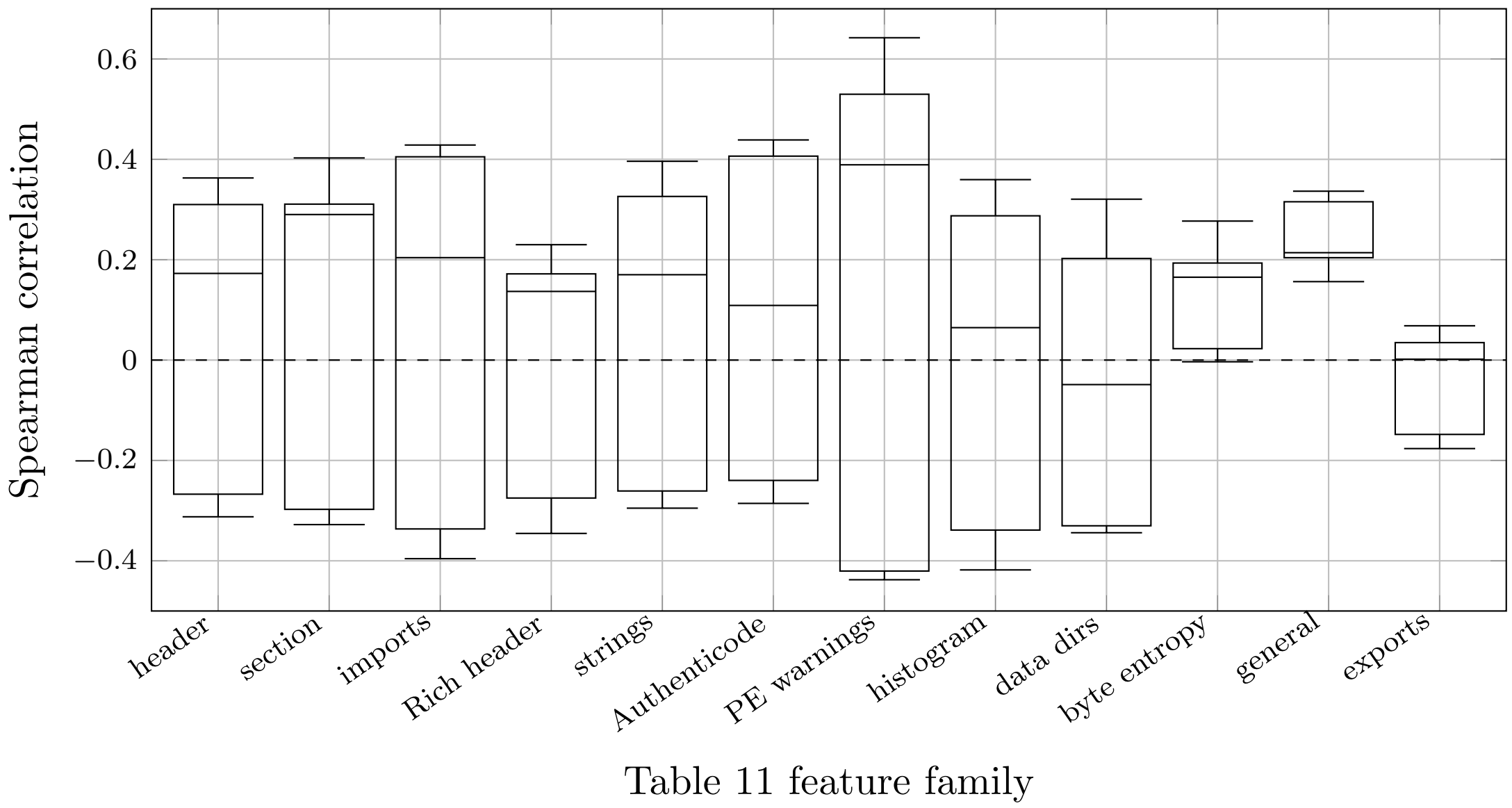
\begin{figure}[H]
\centering
\begin{tikzpicture}
\begin{axis}[
  width=0.95\linewidth,
  height=0.48\linewidth,
  boxplot/draw direction=y,
  ylabel={Spearman correlation},
  xlabel={Table 11 feature family},
  xmin=0.4,
  xmax=12.6,
  ymin=-0.50,
  ymax=0.70,
  xtick={1,...,12},
  xticklabels={header,section,imports,Rich header,strings,Authenticode,PE warnings,histogram,data dirs,byte entropy,general,exports},
  x tick label style={rotate=35, anchor=east},
  grid=major,
  tick label style={font=\scriptsize},
]
\addplot+[black, solid,
  boxplot prepared={draw position=1, lower whisker=-0.312295, lower quartile=-0.267256, median=0.172674, upper quartile=0.309919, upper whisker=0.362860},
] coordinates {};
\addplot+[black, solid,
  boxplot prepared={draw position=2, lower whisker=-0.327853, lower quartile=-0.297451, median=0.290010, upper quartile=0.310723, upper whisker=0.402773},
] coordinates {};
\addplot+[black, solid,
  boxplot prepared={draw position=3, lower whisker=-0.395776, lower quartile=-0.336489, median=0.204106, upper quartile=0.405058, upper whisker=0.428502},
] coordinates {};
\addplot+[black, solid,
  boxplot prepared={draw position=4, lower whisker=-0.345446, lower quartile=-0.274910, median=0.136793, upper quartile=0.171771, upper whisker=0.229918},
] coordinates {};
\addplot+[black, solid,
  boxplot prepared={draw position=5, lower whisker=-0.295106, lower quartile=-0.260886, median=0.170088, upper quartile=0.326057, upper whisker=0.396125},
] coordinates {};
\addplot+[black, solid,
  boxplot prepared={draw position=6, lower whisker=-0.285584, lower quartile=-0.239836, median=0.108879, upper quartile=0.406368, upper whisker=0.438622},
] coordinates {};
\addplot+[black, solid,
  boxplot prepared={draw position=7, lower whisker=-0.437751, lower quartile=-0.420419, median=0.389107, upper quartile=0.529723, upper whisker=0.642258},
] coordinates {};
\addplot+[black, solid,
  boxplot prepared={draw position=8, lower whisker=-0.418138, lower quartile=-0.338797, median=0.064610, upper quartile=0.287471, upper whisker=0.359550},
] coordinates {};
\addplot+[black, solid,
  boxplot prepared={draw position=9, lower whisker=-0.344079, lower quartile=-0.330236, median=-0.048864, upper quartile=0.202465, upper whisker=0.320578},
] coordinates {};
\addplot+[black, solid,
  boxplot prepared={draw position=10, lower whisker=-0.003413, lower quartile=0.022752, median=0.165094, upper quartile=0.193374, upper whisker=0.276976},
] coordinates {};
\addplot+[black, solid,
  boxplot prepared={draw position=11, lower whisker=0.156301, lower quartile=0.204003, median=0.213976, upper quartile=0.315454, upper whisker=0.336511},
] coordinates {};
\addplot+[black, solid,
  boxplot prepared={draw position=12, lower whisker=-0.176373, lower quartile=-0.148155, median=0.001727, upper quartile=0.034782, upper whisker=0.068325},
] coordinates {};
\addplot+[black, dashed, mark=none, forget plot] coordinates {(0.4,0) (12.6,0)};
\end{axis}
\end{tikzpicture}
\caption{Range of Spearman correlations between original EMBER2024 malware probabilities and signed timestamp-artifact family SHAP contributions on the regular test split. Each box summarizes nine values: three original malware models crossed with the all, benign, and malicious row groups.}
\label{fig:original-model-spearman-family-boxplot}
\end{figure}

To test whether the timestamp-artifact signal learned by the obvious-bad classifier is also reflected in the original EMBER2024 malware detectors, we evaluated the downloaded original LightGBM models on the same regular-test PE/.NET rows. The original models use the full 2,568-dimensional EMBER feature vector, so the COFF timestamp column was restored in memory for prediction. For each row, the timestamp-artifact score for a Table~\ref{tab:obviously-bad-family-shap} feature family is the signed family TreeSHAP contribution \(\Phi_{iG}\) from the obvious-bad classifier. For each family \(G\), Spearman correlation was computed between \(\Phi_{iG}\) and the original malware probability within each combination of original model and row group:
\[
  \rho_{m,r,G}
  =
  \operatorname{Spearman}\left(
    \hat{p}^{\,\mathrm{malware}}_{i,m},
    \Phi_{iG}
  \right)_{i\in r},
\]
where \(m\in\{\mathrm{all},\mathrm{PE},\mathrm{type\mbox{-}specific}\}\) and \(r\in\{\mathrm{all},\mathrm{benign},\mathrm{malicious}\}\). Thus each family has nine Spearman values. The distribution of those nine correlations is shown here in Figure~\ref{fig:original-model-spearman-family-boxplot}. As can be seen, the range of correlations is wide and contains zero in all but one case, and so we do not believe the significant benign/malicious information leakage has driven our result. 

\section{Conclusion}
\label{sec:conclusion}

In conducting this research, we found that malware analysts and reverse engineers differed in their views on whether our results were surprising. Still, the exact nature of the results was not anticipated by any domain experts we spoke to. More broadly, our results indicate that this is a very challenging task for current tools. However, current tools have been developed in the context of malware detection rather than compilation-date or first-seen analysis. We believe further work is warranted because, even when first-seen data is available, it is costly and necessarily censored relative to the latent quantities of interest: the ``correct compilation date'' and the ``distribution date.'' Based on our findings, more data is likely needed to yield the most interesting results. This could be more data in the form of longer longitudinal windows, or in the form of libraries/tooling over time to provide potential chronological landmarks. Other methods relevant to this task, such as compilation toolchain identification, could be useful if adapted appropriately. 

\begingroup
\sloppy
\Urlmuskip=0mu plus 2mu
\bibliography{references}
\endgroup

\end{document}